# Earth as an Exoplanet. II. Earth's Time-variable Thermal Emission and Its Atmospheric Seasonality of Bioindicators

Jean-Noël Mettler[1,2] , Sascha P. Quanz[1] , Ravit Helled[2] , Stephanie L. Olson[3] , and Edward W. Schwieterman[4]
[1] ETH Zurich, Institute for Particle Physics and Astrophysics, Wolfgang-Pauli-Strasse 27, CH-8093 Zurich, Switzerland; jmettler@phys.ethz.ch
[2] Center for Theoretical Astrophysics & Cosmology, Institute for Computational Science, University of Zurich, CH-8057 Zurich, Switzerland
[3] Department of Earth, Atmospheric, and Planetary Sciences, Purdue University, West Lafayette, IN 47907, USA
[4] Department of Earth and Planetary Sciences, University of California, Riverside, CA 92521, USA
Received 2022 October 7; revised 2023 February 3; accepted 2023 February 21; published 2023 April 3

## Abstract

We assess the dependence of Earth's disk-integrated mid-infrared thermal emission spectrum on observation geometries and investigate which and how spectral features are impacted by seasonality on Earth. We compiled an exclusive data set containing 2690 disk-integrated thermal emission spectra for four different full-disk observing geometries (North and South Pole-centered and Africa and Pacific-centered equatorial views) over four consecutive years. The spectra were derived from 2378 spectral channels in the wavelength range from 3.75–15.4 $\mu$m (nominal resolution $\approx$1200) and were recorded by the Atmospheric Infrared Sounder on board the Aqua satellite. We learned that there is significant seasonal variability in Earth's thermal emission spectrum, and the strength of spectral features of bioindicators, such as $N_2O$, $CH_4$, $O_3$, and $CO_2$ depends strongly on both season and viewing geometry. In addition, we found a strong spectral degeneracy with respect to the latter two indicating that multi-epoch measurements and time-dependent signals may be required in order to fully characterize planetary environments. Even for Earth and especially for equatorial views, the variations in flux and strength of absorption features in the disk-integrated data are small and typically $\leqslant$ 10%. Disentangling these variations from the noise in future exoplanet observations will be a challenge. However, irrespectively of when the planet will be measured (i.e., day or night or season) the results from mid-infrared observations will remain the same to the zeroth order, which is an advantage over reflected light observations.

*Unified Astronomy Thesaurus concepts:* Astrobiology (74); Earth (planet) (439); Infrared spectroscopy (2285); Biosignatures (2018); Exoplanet atmospheric variability (2020); Space vehicle instruments (1548)

## 1. Introduction

Spatially resolved flyby data from the Pioneer 10/11 (Bender et al. 1974; Baker et al. 1975; Gehrels 1976; Ingersoll et al. 1976; Kliore & Woiceshyn 1976) and Voyager 1/2 (Hanel et al. 1977; Kohlhase & Penzo 1977) missions in the 1970s initiated the exploration of key concepts for the characterization of planetary bodies other than Earth in our solar system. The photometric and spectroscopic observations, ranging from the ultraviolet to infrared (IR), allowed planetary scientists to infer unprecedented details for these worlds such as planetary energy balance, surface, and atmospheric chemical, thermal, and composition properties including cloud and aerosol formation and distribution (for an extensive review see Robinson & Reinhard 2018). In 1993, Sagan et al. (1993) and Drossart et al. (1993) constructed a control experiment by applying remote sensing tools and techniques to search for life on Earth by analyzing Galileo spacecraft (Johnson et al. 1992) Earth-flyby data. Their data indicated a habitable world with water, carbon, and chemical energy. The data also showed signs of biological activity that modulates surface and atmospheric properties. Among these *biosignatures*, the coexistence of $O_2$ and $CH_4$ is a particularly strong indication of life (e.g., Lovelock 1965; Krissansen-Totton et al. 2016, 2018; Schwieterman et al. 2018).

The impact of life on the geochemical environment and the composition of the atmosphere throughout billions of years of coevolution led to the suggestion that alien biospheres should be detectable remotely via spectroscopy (Lovelock 1965; Lovelock et al. 1975; Olson et al. 2018a). Today, the advances made in instrumentation and observing techniques allow us to peak and discover planets beyond our solar system, resulting in a total of 5118 detected exoplanets.[5] Among these discoveries, potentially habitable exoplanets have been found orbiting in the so-called habitable zone (HZ) of their host stars, which sparked interest in spectroscopic studies of exoplanet surfaces and atmospheres for signs of life (e.g., Montet et al. 2015; Anglada-Escudé et al. 2016; Dittmann et al. 2017; Gillon et al. 2017; Gilbert et al. 2020). Thus, over the next decades, the long-run goal of exoplanet science will be the characterization of the atmospheres of temperate terrestrial exoplanets in order to assess their habitability and search for indications of biological activity, which requires the direct detection of their signals over interstellar distances.

The first generation of such terrestrial exoplanet detection and characterization missions will not be capable of spatially resolving the planets due to the large distances of at least several parsecs at which the exoplanets typically will be observed. Even with the most powerful telescopes conceived today, including the recently launched JWST (Gardner et al. 2006), they will remain spatially unresolved point sources. Moreover, the relatively low planet-to-star contrast ratio significantly limits the temporal sampling and the provided [5] http://exoplanet.eu (visited 2022 September 27).

spectral information will be averaged over the observable disk and integration time. The latter may vary between several days and weeks to build up an adequate signal-to-noise ratio to detect biosignatures, depending on the target and mission concept. For example, in the specific case of JWST, which pushes the limits from detecting toward characterizing Jovian to super-Earth exoplanets, the accumulation of transmission spectra from hundreds of transits is required in order to reach a signal-to-noise ratio high enough to potentially confirm the presence of biosignature pairs like $O_2$ and $CH_4$ or $O_3$ and $N_2O$ (e.g., Krissansen-Totton et al. 2016; Fauchez et al. 2019; Lustig-Yaeger et al. 2019; Wunderlich et al. 2019; Tremblay et al. 2020). Hence, considering the mission's lifetime and the telescope time necessary for the detection of atmospheric biosignatures, probably only a few attempts will be made on specific targets. Therefore, JWST as well as other current technologies are not yet capable of detecting and characterizing the atmospheres of temperate, terrestrial exoplanets in a statistically meaningful sample and the community has to wait until space-based direct imaging is realized in future missions like the Habitable Exoplanet Observatory (Gaudi et al. 2020), Large Ultraviolet Optical Infrared Surveyor (Tan et al. 2019) or Large Interferometer For Exoplanets (Quanz et al. 2018).

During the integration time of such direct imaging missions, the spectral appearance and characteristics of a planet change as it rotates around its spin axis and as spatial differences from clear and cloudy regions, contributions from different surface types as well as from different hemispheres evolve with time. In addition, 20 yr of exoplanet discovery have revealed a vast diversity of planets regarding their masses, sizes, and orbits (e.g., Batalha 2014; Burke et al. 2015; Paradise et al. 2022) and it is thought that this diversity also extends to their atmospheric mass and composition, making the characterization of the planetary environment even more difficult. Specifically, the interpretation of the spectrum is not unique and a plethora of solutions exist to describe the planet's surface and atmospheric characteristics.

To achieve the fundamental goal of detecting signs of life on planets beyond our solar system, we will need to be able to interpret this space and time-averaged data. Ideally, an exoplanet candidate with the potential of harboring life would be observed by multiple observing techniques in both the reflected and thermal emission spectrum in order to attempt to fully characterize the planet's nature. Yet, especially for biosignatures, the potential for false positives and false negatives remains (e.g., Selsis 2002; Meadows 2006; Reinhard et al. 2017; Catling et al. 2018; Krissansen-Totton et al. 2022). One way to break this degeneracy and narrow down the set of possible solutions is by adding information coming from time-dependent signals such as atmospheric seasonality.

The phenomenon of planetary seasonality generally arises for nonzero obliquity or orbital eccentricity planets, and the extent of the atmospheric response is governed by stellar flux incident as well as planetary and atmospheric characteristics (e.g., Kopparapu et al. 2013; Guendelman & Kaspi 2019). In our solar system, seasonal variations were observed for the gas giant planets such as Uranus, Saturn, and Jupiter (e.g., Nixon et al. 2010; Fletcher et al. 2015; Shliakhetska & Vidmachenko 2019; Fletcher 2021) as well as for Mars, which is prone to the most diverse seasons in the solar system, due to its 25°.2 tilt of the spin axis and eccentricity of 0.093 (e.g., Leffler et al. 2019; Trainer et al. 2019).

On Earth, the seasonal variation in atmospheric composition, for example of carbon dioxide ($CO_2$), is a well-documented and mechanistically understood biologically modulated occurrence (e.g., Keeling 1960) that is driven by the time-variable insolation and the reacting biosphere. Net fluxes of methane and other trace biological products evolve seasonally, responding to temperature-induced changes in biological rates, gas solubility, precipitation patterns, density stratification, and nutrient recycling (e.g., Khalil & Rasmussen 1983; Olson et al. 2018b; Schwieterman 2018).

Since atmospheric seasonality arises naturally on Earth, it is very likely to occur on other inhabited planets as well. Hence, the search for seasonality as a biosignature on exoplanets is particularly promising and has been proposed by Olson et al. (2018b). Yet, the discussion of time-varying biosignatures has remained qualitative (e.g., Tinetti et al. 2006a, 2006b; Meadows 2006, 2008; Schwieterman et al. 2018) and the field of exoplanet research lacks a comprehensive understanding of which spectral features are impacted by observable seasonality on inhabited worlds and how these impacts would be modulated by stellar, planetary, and biological circumstances.

Earth offers a unique opportunity to study this aspect, yet it requires investigating our planet from a remote vantage point. Although there are several methods to study Earth from afar such as Earth-shine measurements or spacecraft flybys (for a recent review see, e.g., Robinson & Reinhard 2018, and references therein), we chose a remote sensing approach, which offers the extensive temporal, spatial, and spectral coverage needed to investigate the effect of observing geometries on disk-integrated thermal emission spectra and time-varying signals. However, for Earth-orbiting spacecraft it is impossible to view the full disk of Earth and the spatially resolved satellite observations have to be stitched together to a disk-integrated view (e.g., Tinetti et al. 2006a; Hearty et al. 2009; Gómez-Leal et al. 2012).

In a previous paper (Mettler et al. 2020), we analyzed 15 yr of thermal emission Earth observation data for five spatially resolved locations. The data was collected by the Moderate Imaging Spectroradiometer on board the Aqua satellite in the wavelength range of 3.66–14.40 $\mu$m in 16 discrete thermal channels. By constructing data sets with a long time baseline spanning more than a decade and hence several orbital periods, we investigated flux levels and variations as a function of wavelength range and surface type (i.e., climate zone and surface thermal properties) and looked for periodic signals. From the spatially resolved single-surface-type measurements, we found that typically strong absorption bands from $CO_2$ (15 $\mu$m) and $O_3$ (9.65 $\mu$m) are significantly less pronounced and partially absent in data from the polar regions. This implies that estimating correct abundance levels for these molecules might not be representative of the bulk abundances in these viewing geometries. Furthermore, it was shown that the time-resolved thermal emission spectrum encodes information about seasons/planetary obliquity, but the significance depends on the viewing geometry and spectral band considered. In this paper, we expand our analyses from spatially resolved locations to disk-integrated Earth views and present an exclusive data set of 2690 disk-integrated mid-infrared (MIR) thermal emission spectra (3.75–15.4 $\mu$m: $R \approx 1200$) derived from remote sensing observations for four full-disk observing geometries (North and South Pole, Africa and Pacific-centered equatorial view) over four consecutive years at a high temporal resolution (see Figure 1 and Table 1). Using the data set, we assess the dependency of Earth's disk-integrated thermal emission spectrum on observing geometries, phase angles, and integration times much longer than Earth's rotation period as well as investigate which spectral features of habitability and life are impacted by observable seasonality. In Section 2, we describe the input data and our method to derive the disk-integrated spectra, in Section 3 we present and discuss our results.

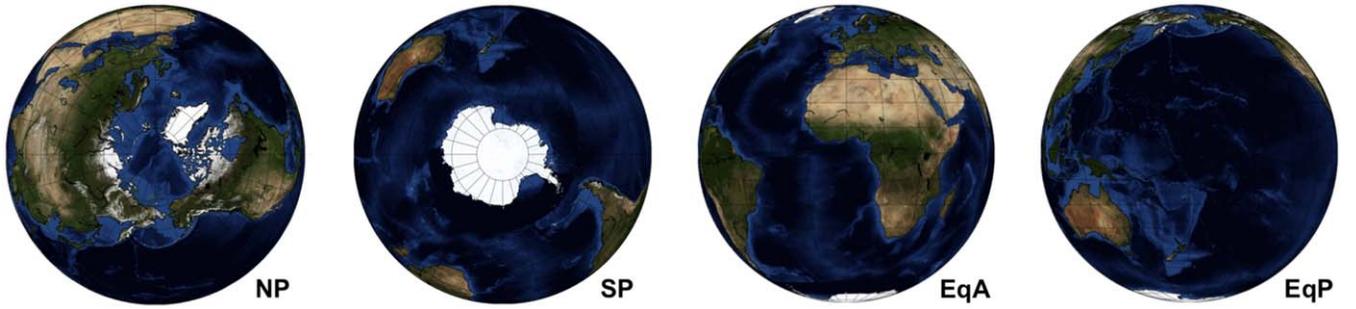

**Figure 1.** The four observing geometries studied in this work. From left to right: North Pole (NP), South Pole (SP), Africa-centered (EqA), and Pacific-centered equatorial view (EqP). In section 3.3, we study integration times longer than the Earth's rotation period. Due to the continuously evolving view of low latitude viewing geometries as the planet rotates, we combine the two equatorial views EqA and EqP to a combined observing geometry, EqC.

In Section 4 we put our findings in context with previous works on this matter and close with the conclusions in Section 5.

## 2. Observations and Data Reduction

We leveraged the extensive temporal, spatial, and spectral coverage of the Atmospheric Infrared Sounder (AIRS) (Chahine et al. 2006) aboard the Earth-monitoring satellite Aqua. Every day, AIRS obtains 2,916,000 Earth spectra in 2378 spectral channels in the MIR wavelength range between 3.75 and 15.4 $\mu$m (nominal resolution: $\lambda / \delta\lambda \approx 1200$). Due to the satellite's operation height of 705 km and due to its Sun-synchronous, near-polar, and circular orbit, it revolves around Earth in 99 minutes, providing a rich set of spectra consisting of day, night, land, and ocean scenes at all latitudes. However, for orbiting spacecraft like this, it is impossible to view the full disk of Earth, which is why the observations have to be tailored to show a spatially resolved, global map of Earth, which can then be disk integrated in order to study Earth's characteristics by means of exoplanet characterization techniques (see Figure 2). For our analysis, we have compiled an exclusive data set of such disk-integrated Earth thermal emission spectra at a high temporal resolution for four different observation geometries over four consecutive years. In total, the data set comprises 2690 spectra.

In order to derive the spectra, we have used radiance measurements from an AIRS IR level 1C product (V6.7) called AIRICRAD (AIRS Science Team/Larrabee Strow 2019),[6] containing calibrated and geolocated radiances given in physical units of W m$^{-2}$ $\mu$m$^{-1}$ sr$^{-1}$ (Manning et al. 2019). These measured AIRS radiances were then mapped onto the globe at high spatial resolution, and subsampled at spatial grid points with Nside = 128 (196,608 pixels) using the Hierarchical Equal Area and Iso-Latitude Pixelization (HEALPIX) approach (Gorski et al. 2005), which allowed us to easily simulate how Earth would look from different perspectives.

**Table 1**
Data Set Overview

| Year | Temporal Resolution | Total Days | Day | Night |
|---|---|---|---|---|
| 2016 | Every 3rd day | 121 | X | X |
| 2017 | Every 3rd day | 121 | X | X |
| 2018 | Daily | 365 | X | ⋯ |
| 2019 | Every 3rd day | 121 | X | X |
| Effective number of spectra | | # | | |
| NP | 405 | | X | X |
| SP | 408 | | X | X |
| EqA_day | 765 | | X | |
| EqA_night | 311 | | | X |
| EqP_day | 391 | | X | |
| EqP_night | 410 | | | X |

The chosen Nside allowed us to sample the data with the best possible resolution. While higher spatial resolution grids would not portray the data correctly, lower Nside value grids resulted in differences in the disk-integrated spectra compared to the full resolution average due to the larger pixel sizes.

For our analysis, we defined four specific observing geometries as shown in Figure 1: North and South Pole, Africa, and Pacific-centered equatorial views. Since Earth-monitoring instruments observe in the nadir viewing geometry, we applied a simple empirical limb correction function adapted from Hodges et al. (2000) to our disk views, where the limb-adjusted radiance, $R(\theta)$, with the zenith angle $\theta$, is calculated from the radiance at nadir, $R(0)$, as follows:

$$R(\theta) = \lambda(\theta) \times R(0),$$

where $\lambda(\theta)$ is the MIR limb correction function as a function of the satellite zenith angle given as

$$\lambda(\theta) = 1 + 0.09 \times \ln(\cos(\theta)).$$

This weighting function progressively down weights off-nadir pixels with their cosine of satellite zenith angle in favor of near-nadir pixels, fully taking account of the geometric effects. Furthermore, due to the swath geometry of satellites, daily remote sensing data contain gores, which are regions with no data points, between orbit passes near the equator. These regions are filled within 48 hr as the satellite continues scanning Earth while orbiting it. However, in order to create snapshots of Earth's full disk on a daily basis, one has to consider the missing data.

For the purpose of investigating the impact of missing thermal emission data on the disk-integrated mean, we analyzed 2000 randomly selected AIRS observation frames from which up to 12% of data was cut out and compared the results of five different interpolation methods (linear, nearest, and cubic python SciPy griddata, nearestND, and python NumPy linear interpolator) to the not interpolated frames. The results showed that the deviation from the original frames to the not interpolated frames with 12% missing data was ≈0.2% and less for the interpolated frames. Thus, the effect of missing thermal emission data on the disk-integrated mean is negligible, if the Earth-view disk contains gores of 12% or less missing data. Hence, due to these results and the fact that AIRS daily coverage is more than 95% of Earth's surface, we have not applied any interpolation methods and refrained from adding artificial data to the scenes. The entire data set can be shared upon request.

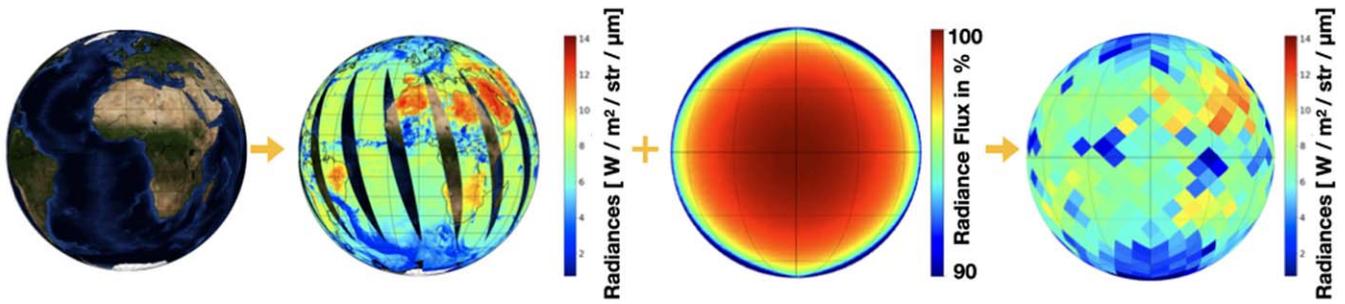

**Figure 2.** An illustration of the method: for every observation geometry the corresponding AIRS radiances were mapped onto the globe and subsampled using the HEALPIX approach. To account for the fact that Earth-monitoring instruments are recording their data in the nadir viewing geometry, a limb-darkening parameterization adapted from Hodges et al. (2000) was applied to the data before disk averaging the disk view. The whole process was repeated for all 2644 MIR thermal emission spectral channels from the AIRICRAD level 1C satellite product, producing the disk-integrated spectra used in this work.

## 3. Results

In the following sections, we analyze the data for the four viewing geometries presented in Figure 1. These viewing geometries evolve throughout the year due to Earth's nonzero obliquity. Figure 3 illustrates how the phases change for the equatorial and pole-on viewing geometries. Whereas the former view blends seasons and has a diurnal cycle, the polar view shows one season but blends day and night. Furthermore, some of the observing geometries discussed in Section 3.3 represent idealized scenarios as they cannot be readily observed for exoplanets by future observatories.

### 3.1. Seasonal Variability of Earth's Thermal Emission Spectrum for Different Viewing Angles

Here, we investigate the annual variability of Earth's MIR thermal emission spectrum due to obliquity as a function of viewing geometry. For the analysis, the measured spectra are considered to be snapshots, i.e., the integration time is a lot less than Earth's rotation period. The results are shown in Figure 4, which displays the time-variable change of flux over one full year for both polar and equatorial viewing geometries. For each specific viewing geometry, the annual average spectrum was calculated from all disk-integrated measurements taken over 4 yr. The plots also show the minimum and maximum measured spectrum within that time period as well as an average summer and winter spectrum. To determine the latter, the months with the highest/lowest flux measurement per year at Earth's peaking wavelength of 10.4 $\mu$m were averaged over 4 yr to get an accurate average spectrum for that season. For the northern hemisphere this turned out to be January and July for the winter and summer seasons, respectively, and vice versa for the southern hemisphere. To facilitate the quantitative analysis of Figure 4, we define the following three atmospheric windows: window 1: 10.2–11 $\mu$m, window 2: 8–9 $\mu$m, and window 3: 3.9–4.1 $\mu$m, which either lie in the IR window (8–14 $\mu$m) or show a maximum absorption of up to ~10%. The results are summarized in Table 2.

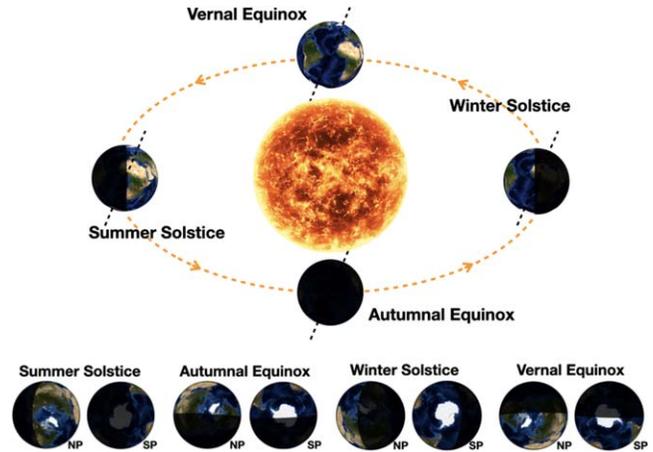

**Figure 3.** Annual change in the position of Earth on its orbit around the Sun. The graphic in the top panel illustrates a view on Earth from a position in space that is at an increased angle with respect to the ecliptic plane in order to show the different phases of illumination during the solstices and equinoxes. The bottom panel shows the phases for the two polar observing geometries at the same time.

The North Pole-centered view (NP), shown in Figure 1, contains a large landmass fraction and latitudes spanning from the arctic circle down to ~20° N. Hence, it comprises three out of the four main climate zones found on Earth, including the arctic, temperate, and tropical zone as well as the subpolar and subtropical transition zones in between them. While the former three climate zones are dominated throughout the year by the same air masses, the subpolar and subtropical transition zones change with seasons as the air masses from neighboring zones enter at various times of the year. This leads, in combination with the surface characteristics of the continental mass, to a larger expected variability. In the NP view, the arctic zone is dominating the scene and contributes therefore the most to the disk-integrated measured flux, followed by the temperate climate zone. The hottest visible climate zone, tropical, is located close to the edge of the scene and its contribution to the overall disk-integrated average is therefore affected by the limb-darkening effect of 5%–10%. The equatorial and sub-equatorial climate zone is not visible for that viewing geometry.

# Earth's Time-Variable Thermal Emission Spectra

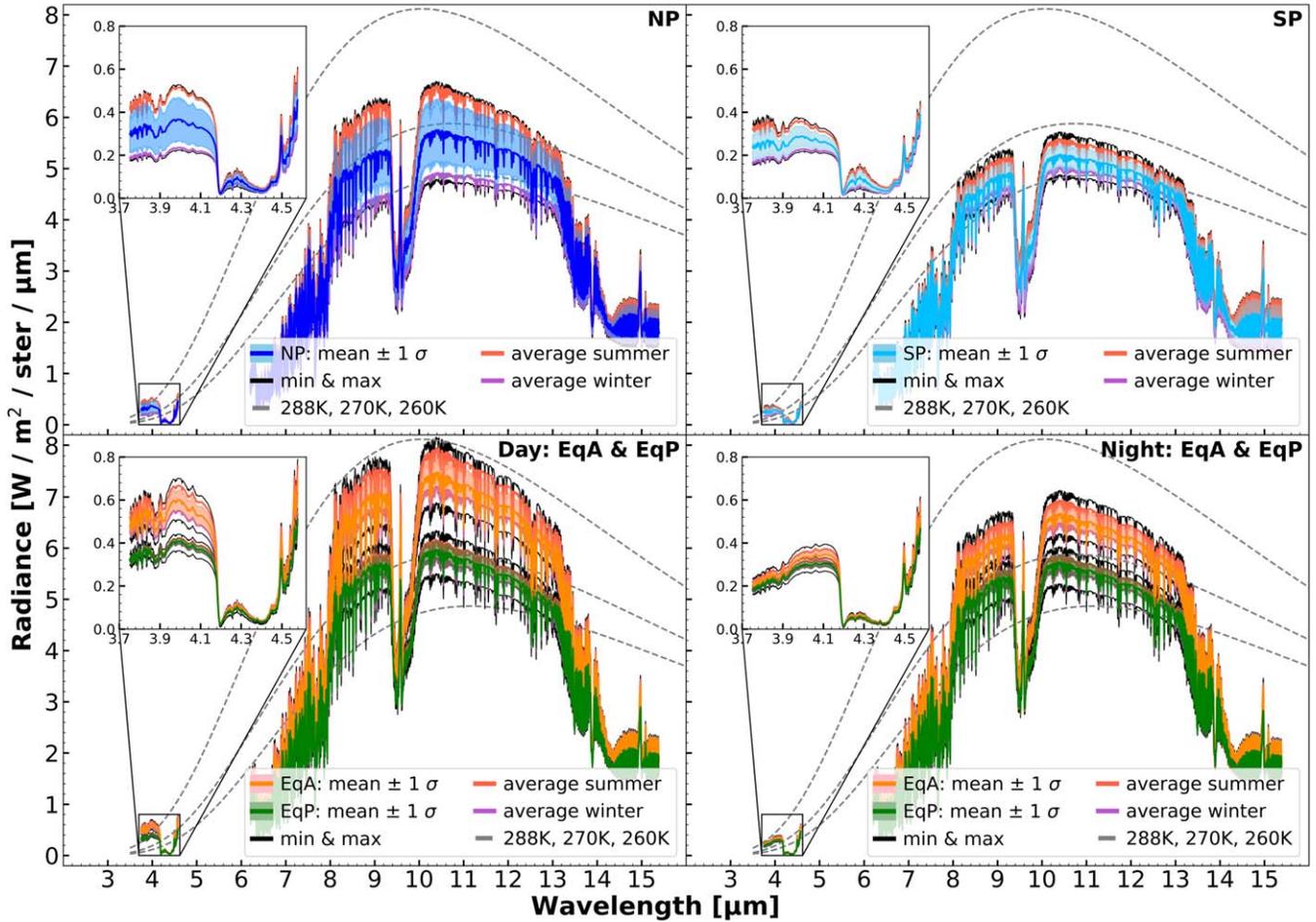

**Figure 4.** A comparison of the disk-integrated thermal emission spectrum for the four different observing geometries (NP, SP, EqA, and EqP). The mean represents the annual spectrum averaged over 4 yr of data. The shaded area corresponds to the standard deviation of all measurements for that particular observing geometry. The average summer and winter were defined as the months with the highest and lowest flux levels at 10.4 $\mu$m, respectively. For the northern hemisphere this turned out to be July and January and vice versa for the southern hemisphere. The dashed lines represent blackbody curves at three different temperatures: 288, 270, and 260 K.

As expected, NP shows the largest seasonal variability out of the four viewing geometries. At longer wavelengths, the flux increases between an average winter and summer by 33% and 42% in the atmospheric windows 1 and 2, respectively. At shorter wavelengths in the MIR (window 3), the relative change in flux increases by more than a factor of 2.

Like NP, the South Pole-centered view (SP) is dominated by the arctic climate zone and the tropical zone lies close to the edge, meaning that the flux coming from this region is affected by the limb-darkening effect. However, due to the inhomogeneous land distribution of our home planet, the pole-on view of the southern hemisphere can be considered as an ocean-dominated view and due to the high thermal inertia of oceans in combination with the dominating arctic climate, the seasonal variability is expected to be less than for the northern hemisphere pole-on view. Comparing the seasonal flux variation of NP to SP shows that the variability is in the order of a third less for longer wavelengths in windows 1 and 2 and a factor of 2 less for shorter wavelengths in window 3. This results in an annual effective temperature change at Earth's peaking wavelength of only 9 K for SP whereas for NP it turns out to be 16 K. In terms of seasonal variability, SP shows a similar variability in the two longer wavelength band atmospheric windows 1 and 2 (11% and 13%, respectively) than the Pacific-centered, ocean-dominated, equatorial view EqP view. However, at shorter wavelengths, e.g., window 3, SP shows a 15% larger variability than EqP. The increased variability at the shorter wavelengths in window 3 could be assigned to a reflected light component and the difference in landmass fraction and its surface characteristics contained in these two observing geometries. The pole-on view of the southern hemisphere is centered on the antarctic continent Antarctica and its ice shelves.

The dominating climate zones for the Africa-centered equatorial viewing geometry (EqA) are the equatorial and tropical zones as well as the subequatorial and subtropical transition zones linking them. The EqA view includes an extended equatorial zone as both the northern part of South America as well as central Africa lie in the view, yet, radiances from the latter contribute more to the disk-integrated mean. Fluxes coming from the temperate, and especially, the arctic climate zones are progressively down-weighted by the limb-darkening parameterization. As expected, the EqA view shows the highest flux readings of all the viewing geometries during both day and night time in the summer season, reaching an effective temperature of 288 K at Earth's peaking wavelength.

**Table 2**
Seasonal Flux Variability—the Figures Represent the Relative Change in Flux of an Average Summer versus Winter for a Specific Viewing Geometry and Have Been Rounded to the Nearest Integer

| Observing Geometries | Window 1 (10.2–11 $\mu$m) | Window 2 (8–9 $\mu$m) | Window 3 (3.9–4.1 $\mu$m) |
|---|---|---|---|
| Summer vs. winter | | | |
| NP | 33% (39%) | 42% (49%) | 118% (135%) |
| EqA (day + night) | 22% (29%) | 26% (35%) | 105% (130%) |
| EqP (day + night) | 11% (19%) | 12% (22%) | 45% (65%) |
| SP | 11% (17%) | 13% (20%) | 60% (70%) |
| EqA (day) | 10% (18%) | 12% (20%) | 20% (39%) |
| EqA (night) | 7% (13%) | 8% (15%) | 12% (30%) |
| EqP (day) | 6% (15%) | 7% (17%) | 7% (20%) |
| EqP (night) | 5% (13%) | 5% (15%) | 9% (25%) |
| Day vs. night | | | |
| EqA summer | 14% (14%) | 17% (16%) | 77% (75%) |
| EqA winter | 11% (11%) | 13% (9%) | 65% (65%) |
| EqP summer | 5% (5%) | 6% (6%) | 35% (35%) |
| EqP winter | 3% (3%) | 4% (4%) | 35% (35%) |

**Note.** The percentages given in the brackets indicate the relative change in flux of the range between the maximum and minimum measured spectrum. Since the data sets of NP and SP naturally include day (summer) and night (winter) data, the variability for the EqA and EqP views (day + night) was determined using an average summer and winter from the day and night data set, respectively. In the lower third of the table, we compare day versus night spectra for the denoted season. To facilitate the quantitative analysis of Figure 4, three spectral windows were defined that lie in atmospheric windows.

Its seasonal variability in thermal emission is 22%–26% in windows 1 and 2, respectively, the second highest after NP although at shorter wavelengths. The relative change in flux of the range of measured spectra is reaching a similar value (130%). The additional variability and spread in data could be due to the influence of clouds, lowering the overall average of the scene as the contrast between hot and cold regions within the same view is higher than for viewing geometries with a dominating cold region like the arctic climate zone in the case of NP. The relative change in flux between summer and winter at day times (night) is 10% and 12 % (7% and 8%) in windows 1 and 2, respectively. At shorter wavelengths, the difference is a factor of 2 larger for the day data, while for the night it is only 5% larger. However, when the whole range of measured spectra is considered, the night data also shows a thermal emission variability of a factor of 2 larger at shorter than at longer wavelengths. The flux variability due to day and night at longer wavelengths is very similar for the summer and winter seasons, deviating only by 3%–4% in windows 1 and 2, resulting in a change of effective temperature of 7–8 K. Yet, the diurnal cycle has a larger impact on fluxes measured at shorter wavelengths where the flux readings vary by a factor of 4.5 and 5.5 for summer and winter, respectively, compared to the longer wavelength regime. A contributing cause to the increased variability in the day is the reflected light component of the radiation in window 3 which is larger for the continental areas versus the ocean-dominated areas.

The second equatorial viewing geometry studied in this work is Pacific-centered, ocean-dominated, equatorial view (EqP) and comprises the same climate zones as the EqA view.

However, due to the lack of landmass near the Equator, the equatorial climate zone is not as prominent and extended as in the EqA view, making the tropical and subtropical climate zones the dominating climates for that particular Earth view. Furthermore, EqP shows the largest ocean-mass fraction from all the observing geometries; hence, it is expected to show the least variability in thermal emission because of the large thermal inertia of oceans. The subarctic and arctic climate zones as well as the majority of the visible landmass are located close to the edge, where the fluxes are affected by the limb-darkening effect. The largest contribution from a landmass is coming from the Australian continent. EqP's seasonal thermal emission variability between an average summer day and an average winter night is around 11% for the longer wavelengths and 45% in window 3. Comparing these values to the EqA view, which includes the same climate zones but has a different landmass fraction, the variability is a factor of 2 and less in all three windows and even a factor of 3 less when it is compared to the NP viewing geometry. In terms of seasonal variability, the EqP view is similar to the SP viewing geometry, showing that the dominant factor in keeping the thermal emission variability low is the high thermal inertia of the oceans. Moreover, if Earth was observed in the EqP perspective there is no benefit whether the planet is observed during day or night time or in which season since the increase in thermal emission flux is negligibly small (between 3% and 7%) at longer wavelengths. At shorter wavelengths, there is 35% more flux in window 3, independent of the summer or winter seasons.

### 3.2. Earth's Degeneracy in Thermal Emission Spectra

When we observe exoplanets, we will not have prior knowledge of when or from which direction we are attempting to characterize it. We simulated this uncertainty by viewing Earth randomly and found strong spectral degeneracy with respect to viewing geometry. This degeneracy arises due to the variable blending of time-variable thermal emission from hemispheres with opposing seasonal signals in disk-integrated views. This complexity complicates remote characterization of planetary environments. Specifically, the interpretation of the spectrum is not unique and a plethora of solutions exist to describe the planet's surface and atmospheric characteristics, which, in the context of exoplanets, imposes another challenge, especially for planets whose orbital elements and obliquity are not well constrained.

Three out of the four observing geometries that we have studied for Earth in this work cannot be distinguished from each other with single-epoch observations. The large annual variability of the northern hemisphere NP viewing geometry covers the flux range of EqP and SP and is exceeding lower flux readings than the southern hemisphere view SP for an average winter. Differentiating between the two hemispheres is especially challenging during March (vernal equinox) where the spectra of NP and SP overlap in all three spectral windows. The same is the case in November with overlapping spectra in the longer wavelength range windows 1 and 2. However, at shorter wavelengths, especially between 3.75 and 4.1 $\mu$m, there is a gap of ~40% flux difference between SP to NP indicating the different seasons the hemispheres are in at that moment of the orbit as well as including the flux difference in day versus nightside. In general, the northern hemisphere pole-on view emits 20%–30% more flux at longer wavelengths (windows 1 and 2) and between 50% and 150% at shorter wavelengths (window 1) during the summer solstice and autumnal equinox than the southern pole-on view SP (see Figure 5). During the winter solstice, summer on the southern hemisphere, SP emits ~10% more flux at the longer wavelengths and 20%–40% at shorter wavelengths than the northern hemisphere NP view. The increased variability in flux at shorter wavelengths in the day is due to a contributing factor of a reflected light component.

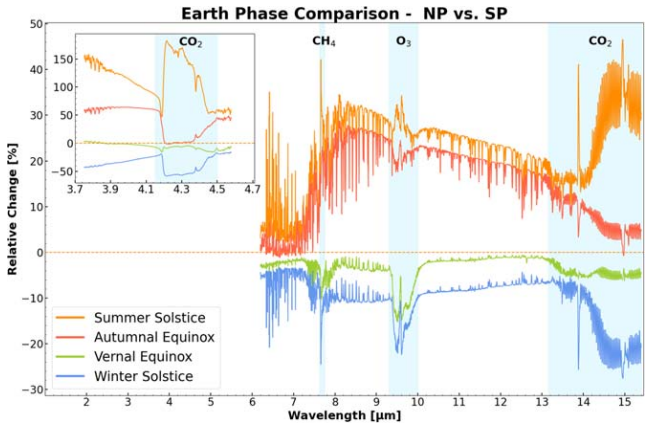

**Figure 5.** Comparing the spectra of the northern to the southern hemisphere at the solstices and equinoxes that occur in June and December and March and September, respectively. The figure shows the relative change in flux, on the *y*-axis, as a function of wavelength. For this analysis, monthly averaged spectra for these months were calculated.

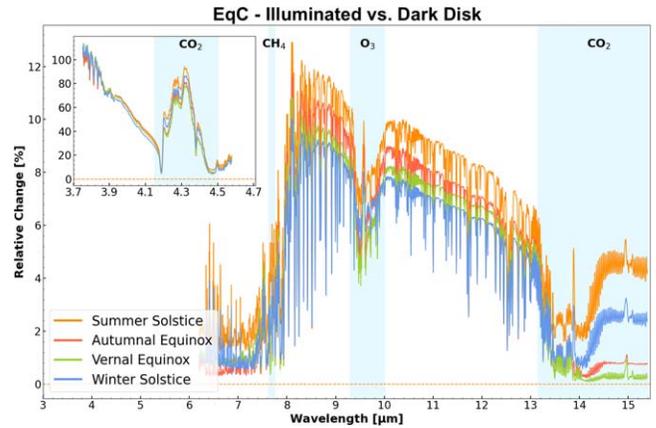

**Figure 6.** Comparing the day vs. nightside (illuminated vs. dark disk) at different times on Earth's orbit for the combined equatorial view EqC.

Moreover, over the course of a year, NP reaches similar flux readings at longer wavelength ranges as the Pacific-centered view EqP in May and September (autumnal equinox). Although there is a gap at shorter wavelengths between the two May spectra, during the autumnal equinox, it is hard to differentiate between a pole-on northern view to an ocean-dominated equatorial view. The southern hemisphere SP and EqP show a similar annual variability at longer wavelengths. Differentiating between these two observing geometries is particularly difficult for a quarter of the year as their spectra overlap between December through February when the southern hemisphere's summer is entering the lower flux range of the equatorial winter. At shorter wavelengths, the spectra of SP only overlap with EqP at the peak of its summer in December.

Although the Africa-centered observing geometry EqA shows the second largest annual variability, the flux level range in atmospheric windows 1 and 2 remains well separated from other viewing geometries and is therefore the only viewing geometry whose range is not intersecting with others. At shorter wavelengths, however, spectra from July for the NP viewing geometry reach flux levels equivalent to EqA's spectra during winter time.

Hence, without sufficient knowledge about a planet's orbital parameters and obliquity, interpreting the space and time-averaged data based on single-epoch measurements is quite challenging—even for Earth. Therefore multi-epoch measurements and the resulting time-dependent signals may be required to help break the degeneracy in the thermal emission spectra. As previous work has shown, different surface types have different photometric and thermal properties and by adding the time factor, i.e., variability, additional information can be gained for assessing a planet's characteristics and habitability (e.g., Tinetti et al. 2006b; Gómez-Leal et al. 2012; Madden & Kaltenegger 2020; Mettler et al. 2020; Lehmer et al. 2021).

### 3.3. Observing Earth from an Equatorial Viewing Perspective

In this section, we focus on the equatorial viewing geometry and study Earth's thermal emission spectrum as a function of seasons and phase angles, i.e., different contributions of the day- and nightside from the planet, at four specific points in time of Earth's orbit: summer and winter solstice and the vernal and autumnal equinox, as shown in the top panel of Figure 3. The solstices and equinoxes occur in June and December and March and September, respectively.

For the analysis, we calculate the monthly mean spectra averaged over 4 yr of the aforementioned months. Hence, we are considering integration times much longer than Earth's rotation period and therefore combine the two equatorial views, EqA and EqP, and denote the resulting data set EqC. The day and night spectra of the new data set and how it compares to the polar views are shown in Figure 11 in Appendix B.

In order to compose the full- and new-Earth phases, day and night equatorial data sets have been used, respectively, while for the half-Earth waxing and waning phase, 50% day and 50% night data of the corresponding months was used. Figure 6 shows the comparison of the relative change of flux between the day versus nightside spectrum at a given location on Earth's orbit.

In the wavelength region between 8 and 9 $\mu$m and 10 and 11 $\mu$m the flux levels from the dayside are between 9% and 12% and 8% and 10% larger, respectively. The day versus nightside difference varies by ~2.5% for the seasons and these band regions, where the largest relative change is shown during summer followed by fall, spring, and winter in descending order. The order, however, rearranges in wavelength regions with absorption features. At the band centers of $CO_2$ and $O_3$ for example, the difference between day and nightside at winter solstice shows the second largest relative change. At the center of the 15 $\mu$m $CO_2$ feature, Earth emits ~2.5% and ~4.5% more flux from the fully illuminated disk for the winter and summer solstice, respectively, which is 5–11 times more than at the autumnal and vernal equinoxes. Similarly to the 15 $\mu$m feature, the $CO_2$ feature centered at 4.3 $\mu$m shows the same behavior except that the difference in emission due to the day- and nightside is as large as ~70%–85%. In general, at shorter wavelengths, 3.7–4.6 $\mu$m, the difference between the fully illuminated and dark disk is the strongest, reaching levels above 100%, indicating a significant contribution from reflected light.

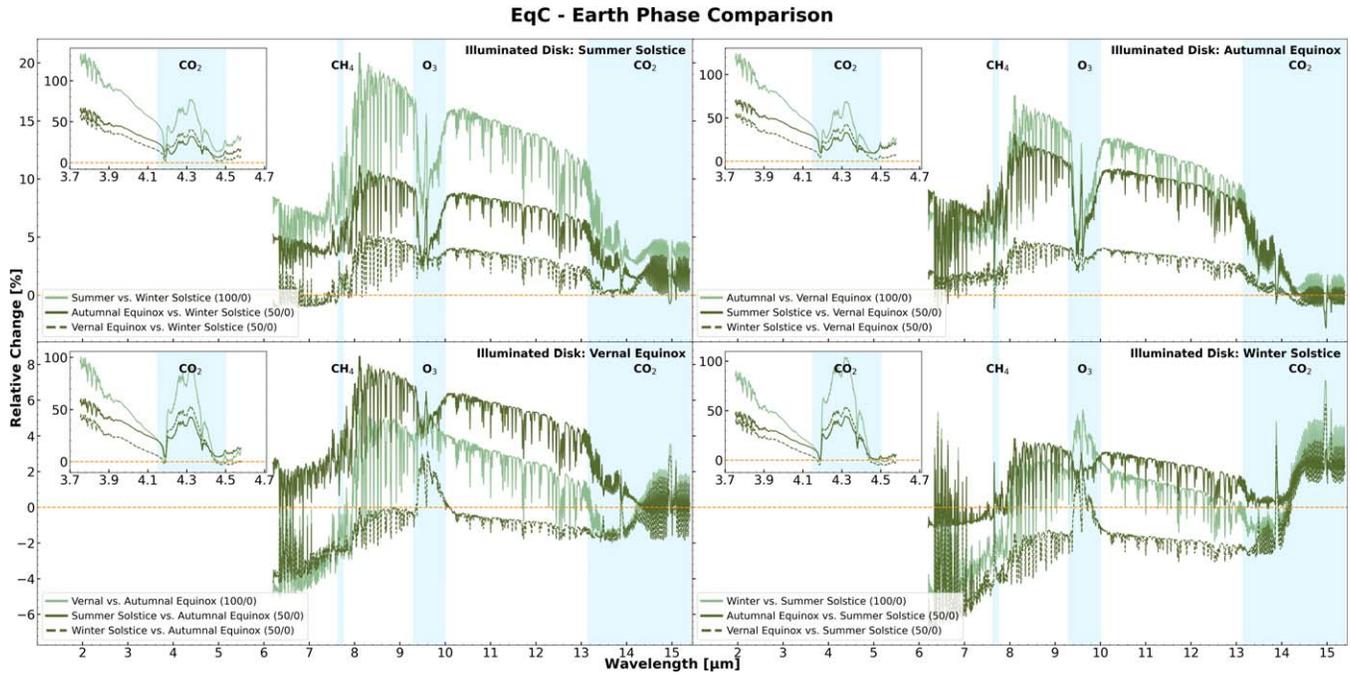

**Figure 7.** Comparing how Earth's spectral appearance changes along one full orbit, given a constant equatorial viewing geometry. Each panel corresponds to a viewing geometry for a remote observer as shown in Figure 3 with the relative change in flux on the y-axis and the wavelength range on the x-axis. The colors as well as the percentages given in the brackets indicate the current phase type of the target position, where 100, 50, and 0 correspond to the illuminated, partially illuminated, and dark disk, respectively. Earth's position during the secondary eclipse for that specific viewing geometry is given in the top right.

The results of the analysis of Earth's thermal emission spectrum as a function of seasons and phase angles are shown in Figure 7. Each panel corresponds to a constant equatorial viewing geometry as seen by a remote observer. An example is illustrated in Figure 3. The observation direction is indicated by Earth's position during the secondary eclipse in the top right corner of the panels. Keeping the observation direction constant, the panel compares how Earth's spectral appearance changes along one full orbit. Specifically, the relative change in flux is calculated between the target position and winter solstice in order to compare the thermal emission spectra of the different positions on Earth's orbit to each other at increments of a quarter orbit. Note that our analysis corresponds to ideal conditions in comparison to what actual observations would be. Scenarios where the planet is behind or in front of its star are not readily observable.

Depending on the observation direction, Earth's phase at winter solstice changes. Thus, for example, the top left panel in Figure 7 corresponds to what a remote observer looking at the Earth–Sun system as shown in Figure 3 from the right-hand side would measure. During summer solstice Earth's disk is fully illuminated and partially illuminated at the vernal and autumnal equinox. At the winter solstice, Earth is in its new-Earth phase. In this perspective, for a remote observer the largest contrast in the relative change in flux would be when the Earth was observed during winter and half an orbit later during the summer solstice, where Earth emits 11%–19% more flux between 8 and 13 μm. The window region between 8 and 12 μm is especially important since it offers the best chance to estimate a planet's surface temperature (e.g., Des Marais et al. 2002). For the same window region, the difference in flux between springtime (vernal equinox) and winter solstice lies between ∼2.5% and 5% and between 6% and 10% during the autumnal equinox. Hence, from the four panels in Figure 7 it is apparent that remote observers detecting Earth in an equatorial viewing geometry, where the primary eclipse occurs at the winter solstice and the secondary eclipse around the summer solstice (from the right-hand side in Figure 3) have the most favorable viewing angle and phase configurations for characterizing seasonal differences in the flux of Earth.

The second best equatorial viewing geometry for detecting our planet's seasonal changes in flux is for observers looking from a direction where Earth's secondary eclipse would occur around the autumnal equinox. In the window region between 8 and 12 μm, the fully illuminated disk around the autumnal equinox emits still between 10% and 16% more flux than at the vernal equinox where Earth's nightside is facing the observer. Furthermore, although being in the half-Earth waxing phase, around summer solstice Earth emits ∼2% less flux than the fully illuminated disk due to the hotter temperature and the rather large landmass of the African continent. The relative change in flux of the winter solstice versus the vernal equinox is up to ≈5%, which is similar to the top left of Figure 7 described above. This is mainly due to similar temperatures and the large ocean fraction in the equatorial viewing geometry for Earth, which is rather resistant to temperature changes.

For an observing geometry where the Earth's disk is fully illuminated around the winter solstice (from the left-hand side in Figure 3), Earth's nightside is facing the observer during the summer solstice. Compared to the vernal equinox waning phase, Earth's nightside emits ∼2% more flux between 8 and 12 μm except in the ozone absorption feature. Comparing Earth's emission from the autumnal equinox and winter solstice to the summer solstice nightside, they emit between ∼1% and 3.5% more flux. Hence, for such a viewing geometry, determining whether the planet has a nonzero obliquity, and thus, seasons would be very challenging.

### 3.4. Observable Seasonality on Earth

The change in insolation, due to Earth's axial tilt, causes the photometric and spectroscopic appearance of our planet to change on hourly, seasonal, and annual timescales. The inhomogeneous distribution of net radiation imbalances drives the global circulations of the atmosphere as well as oceans, and although Earth's eccentricity is nearly circular, there is evidence of greater irradiance in the southern summer than the northern latitudes in the northern summer. Winds and weather patterns adjust to transport heat from higher irradiated regions to lower ones, introducing climatic variability. While the change in insolation impacts photochemical processes at higher altitudes, affecting the atmospheric composition as well as its vertical temperature structure, the biosphere modulates the seasonal atmospheric composition. Net fluxes of methane and other trace biological products evolve seasonally, responding to temperature-induced changes in biological rates, gas solubility, precipitation patterns, density stratification, and nutrient recycling (e.g., Khalil & Rasmussen 1983; Olson et al. 2018b). Thus, these temporal modulations can take the form of oscillations in gas concentrations or surface spectral albedo.

Since atmospheric seasonality arises naturally on Earth, it is very likely to occur on other nonzero obliquity and eccentricity planets, and if they are inhabited, life may modulate the seasonal variations in atmospheric composition as well (e.g., Olson et al. 2018b). However, these time-dependent modulations, of physical or biogenic origin, must be present and observable in the disk-averaged spectra of those planets if we hope to leverage these signals to recognize exoplanet life. As discussed in the literature (e.g., Ford et al. 2001; Hearty et al. 2009; Gómez-Leal et al. 2012; Schwieterman et al. 2018) as well as in Sections 3.1–3.3 in this work, there is a strong seasonal and viewing geometry dependency, posing observational challenges to detect seasonal signals that are driven by obliquity in addition to the current technical limitations of detecting and characterizing terrestrial HZ planets. Seasonal contrast increases with obliquity and the effect of obliquity is the strongest near the poles. Yet, observing in a near pole-on view will probe only one hemisphere, while an equatorial view includes variabilities of both visible hemispheres if hemispherical asymmetries, for example in landmass distribution, exist that generate the obliquity-driven seasonality. Depending on the landmass fraction and distribution as well as orbital and viewing geometry configurations, the observed magnitude of seasonality will be muted. This relationship arises due to latitude averaging, including contributions from opposing seasons in each hemisphere, in disk-integrated spectra.

In this section, we leverage the temporal and spectral coverage of our data set in order to quantify the time-varying signal of Earth's observable biosignatures like methane, ozone, and nitrous oxide in the MIR and address the question of whether the variability is detectable to a remote observer. In addition, we investigate the seasonality of carbon dioxide due to its role in climate regulation and biological activity such as photosynthesis, for example.

#### 3.4.1. Calculating Equivalent Widths

For the quantitative analysis, we calculate the equivalent width (EW), a measure of the strength of an absorption feature, which is defined by

$$W_\lambda = \int \left(1 - \frac{I_{CRS}}{I_c}\right) d\lambda \qquad (1)$$

where $W_\lambda$ is the EW and $I_{CRS}$ and $I_c$ the radiance from the continuum removed spectrum and continuum, respectively. In order to calculate the EWs, for each disk-integrated spectrum, we calculated a continuum removed spectrum (CRS) by applying a convex hull technique where we subtracted the difference between the hull and the original spectrum from a constant. Thus, absorption features within a CRS are always between zero and one and can be easily identified by the points where the spectrum is touched by its convex hull, giving the full depth of the feature. In order to isolate the molecular absorption features of $O_3$, $CH_4$, $CO_2$, and $N_2O$, we have adopted the spectral band centers and intervals from Catling et al. 2018 and evaluated the normalized radiances at these fixed bandwidth ranges. While the full feature depth of $O_3$ centered at 9.6 $\mu$m and $CO_2$ centered at 4.3 $\mu$m can be inferred directly, the molecular absorption features of $N_2O$ and $CH_4$ in the MIR at 4.5 and 7.7 $\mu$m, respectively, lie within an $H_2O$ feature with the center at 6.3 $\mu$m. To exclude additional variation induced by the $H_2O$ feature in the isolated features of $N_2O$ and $CH_4$, we have defined the shoulder points to be at the limits of their bandwidths. Figure 8 displays an example of the four isolated absorption features with the normalized radiances on the *y*-axis and wavelength range on the *x*-axis. The convex hull or the baseline is shown in red.

#### 3.4.2. Physical Conditions That Affect EWs

Besides chemistry, the dominant mode of interaction between molecules and MIR radiation in the troposphere and stratosphere is absorption. IR-absorbing molecules absorb at a wide range of wavelengths corresponding to transitions between different forms of energy levels (rotational and vibrational). The corresponding cross sections vary by many orders of magnitude, depending on the nature of the molecule and the transition. Most of the absorption features found in the MIR spectrum are vibration-rotation bands, for which the band center is determined by a vibrational transition, with simultaneous rotational transitions forming branches of discrete lines on either side of the center (e.g., Goody & Hu 2003). Line and band strengths, and thus the EW, are proportional to molecular number density, absorption coefficient per molecule, optical path, and lower state population, which is governed by Boltzmann's law and can be highly temperature dependent. While line absorptions saturate with increasing path length, continua, e.g., the water vapor continuum, behave differently and do not saturate. Their relative importance increases at long absorption paths.

Physical conditions in planetary atmospheres can lead to different types and degrees of line broadening that can affect EWs. For Earth and the observational data presented in this work, relevant line-broadening mechanisms are pressure (collisional) broadening and thermal (Doppler) broadening. The former is the dominant broadening mechanism in the lower atmosphere (troposphere) and is due to collisions between chemical species with the collision frequency being a strong function of pressure. The thermal broadening is caused by the line-of-sight thermal velocity distribution of molecules at a given temperature in the planetary atmosphere (e.g., Hedges & Madhusudhan 2016). This type of broadening becomes the dominant mechanism above the Tropopause at $\approx$10 mbar 30 $km^{-1}$ (e.g., Goody & Hu 2003). Furthermore, water vapor is a very efficient agent to broaden spectral lines of $CO_2$, $N_2O$, $CH_4$, and other gases. The broadening by water vapor is much larger than that of nitrogen and oxygen, which are the two main contributors to dry air broadening (e.g., Tan et al. 2019). The amount of water vapor in the terrestrial atmosphere is highly variable both spatially and temporally and can account for up to 5% of the atmosphere in the tropics. Thus, the broadening due to the presence of water vapor as well as the temporal variability in its abundance affects the EW measurement and induces additional variability to the atmospheric seasonality of the molecular absorption features studied in Section 3.4.3.

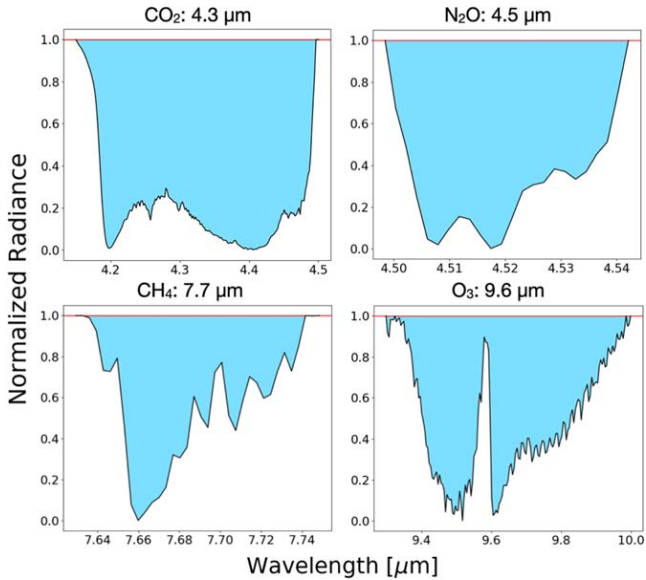

**Figure 8.** The four isolated absorption features of $CO_2$ (4.3 $\mu$m), $N_2O$ (4.5 $\mu$m), $CH_4$ (7.7 $\mu$m), and $O_3$ (9.6 $\mu$m) with the normalized radiance and the wavelength on the $y$-axis and $x$-axis, respectively. In order to isolate these spectral features, we have applied a hull difference continuum removal technique, where a convex hull was fitted over the spectrum to define the continuum. The hull difference is obtained by subtracting the difference between the hull and the original input spectrum from a constant. The EWs were then calculated according to Equation 1. The bandwidths of the molecular absorption features are 350, 40, 120, and 700 nm for $CO_2$, $N_2O$, $CH_4$, and $O_3$, respectively.

The EW measurements are also affected by nonlocal thermodynamic equilibrium (nLTE) effects that have a large emission contribution at the core of a spectral band. nLTE emission occurs generally above the stratopause, where solar pumping populates the vibration-rotation energy levels more quickly than collisions can thermally redistribute the energy (e.g., DeSouza-Machado et al. 2007). While nLTE emission is negligible at 15 $\mu$m, all stratospheric as well as many upper atmosphere bands like the 4.3 $\mu$m spectral region are impacted.

In addition, changes in the atmospheric temperature structure also affect the EW measurements. The observed emission peaks (e.g., shown in Figures 4 and 8) at 9.6 and 15 $\mu$m are caused by layers that are warmer than the top of the troposphere. Because of the high emissivities corresponding to these wavelengths, the radiation escapes into the stratosphere and is emitted only when the partial pressure of $CO_2$ at 15 $\mu$m or $O_3$ at 9.6 $\mu$m is sufficiently low to no longer absorb at these wavelengths. Since the stratosphere gets warmer with increasing altitude, the emission originates from a warmer layer than the top of the troposphere. Thus, changes in the atmospheric temperature structure alter the emission peaks and therefore affects the EW of the absorption feature.

### 3.4.3. Analyzing the Time Series of EWs

Figure 9 displays the 4 yr average of the annual change in EWs for each viewing geometry and bioindicator. Day zero represents January 1. The four molecular absorption features investigated in this work show an annual variation in strength. As expected, the largest amplitudes are shown by the two polar views, NP and SP. Furthermore, the variation time series of the northern hemisphere is opposite to the southern one due to the seasonal cycle of solar insolation. The results are summarized in Table 3 and the seasonality plots for the individual viewing geometries are attached in Figures 12–15.

Ozone ($O_3$) reaches peak concentrations of up to 10 ppm in the stratosphere between 15 and 30 km in altitude on Earth and is the result of photochemical reactions that split oxygen (e.g., Grenfell et al. 2007, Figure 1). Yet, the abundance as well as the altitude of the peak vary spatially throughout the year and hemispheres. Since the feature at 9.6 $\mu$m is highly saturated, its strength is expected to essentially remain unchanged. From Table 3, we deduce the global annual mean EW for $O_3$ which deviates by 1.03% for the different observing geometries. While the equatorial regions deviate by 0.15% ($\sigma = 0.72$ nm) from their mean, the largest difference contribution to the global mean can be assigned to the polar regions whose mean EWs differ by 1.05% ($\sigma = 3.76$ nm). The latter also show oscillations with amplitudes of 6.68 and 4.02 nm for NP and SP, respectively, which are a factor of $\sim$2–3 larger than for the equatorial views (EqA: 2.65 nm and EqP: 2.14 nm). The amplitudinal difference for the atmospheric seasonality of ozone results in a 40% difference between the two hemispheres in favor of NP, while the EqA view shows a 19% larger seasonality in the strength of the $O_3$ feature than the ocean-dominated view EqP. Comparing their seasonal range to the annual mean EWs of the ozone absorption feature results in an annual change in strength, or seasonal variation, of 3.72% and 2.27% for NP and SP and 1.51% and 1.22% for the EqA and EqP views, respectively.

Although the band strength in the IR is rather sensitive to the temperature differences between the lower and middle atmosphere, the ozone spectral feature can be weakened by clouds (e.g., Kitzmann et al. 2011). Cross checking with level 3 satellite data (AIRS3STD (AIRS Science Team/Joao Teixeira 2013)), Figure 10 panel (e) shows that the $O_3$ abundance trend follows the oscillations shown in the first panel of Figure 9 for all the viewing geometries. Even though this might suggest that the EW oscillation is mainly due to variations in $O_3$ abundance, we cannot exclude contributions from changes in the stratospheric temperature and variability in patchy cloud coverage that decrease the emitted continuum flux and reduce the relative depths of spectral features (e.g., Des Marais et al. 2002).

Methane ($CH_4$) is the seventh most abundant atmospheric constituent in modern Earth's atmosphere with a surface concentration of 1.87 ppm (Canadell et al. 2021). Roughly 90% of the net surface source is associated with the respiration of methanogenic microbes. Its presence in the atmosphere in combination with $CO_2$ alongside the absence or low abundance

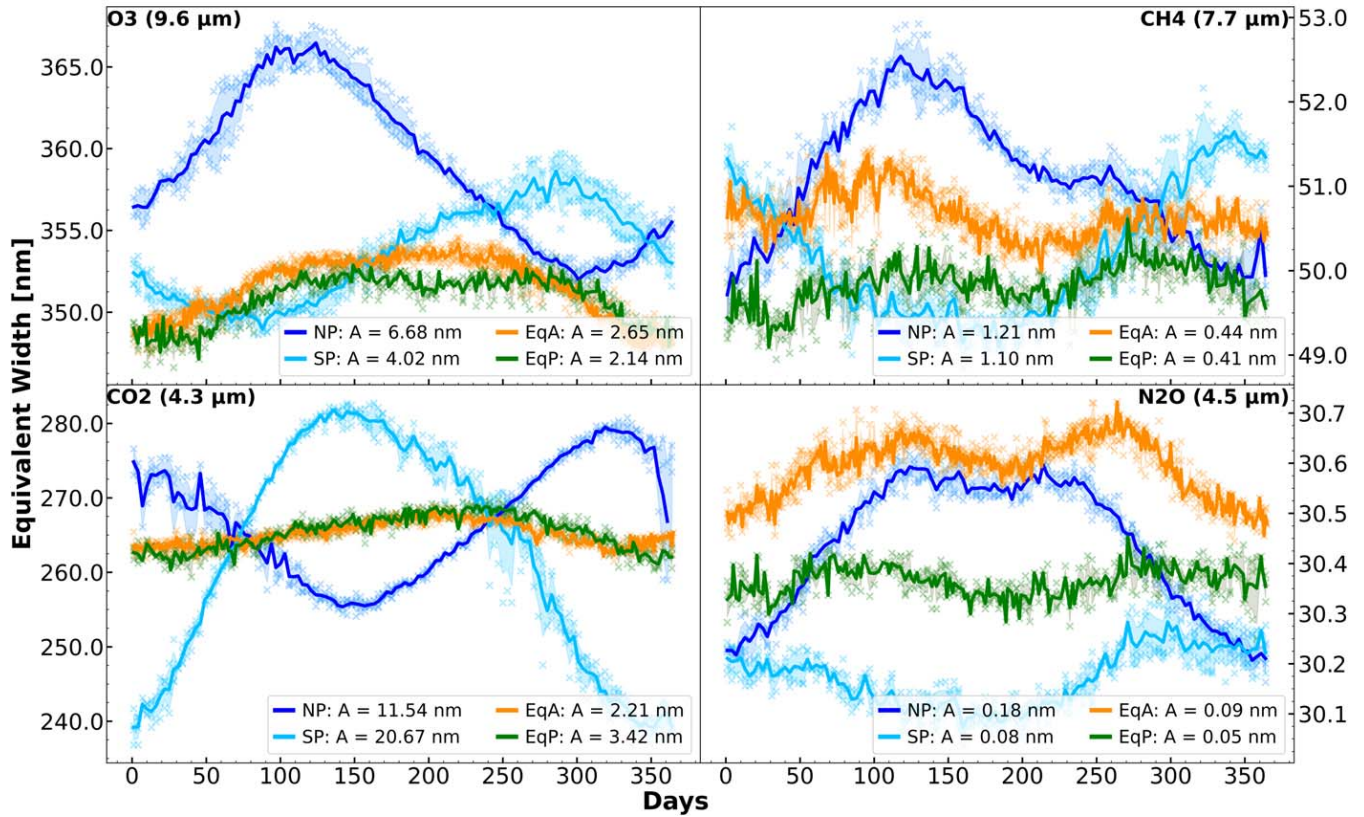

**Figure 9.** Atmospheric seasonality: 4 yr average of the time-varying EW signal for the different viewing geometries and bioindicators. The EWs in nanometers and the days are given on the *y*-axis and *x*-axis, respectively. Day zero represents January 1. The solid lines and shaded areas represent the 4 yr average and 1 standard deviation of all EW measurements, which are shown as a scatter plot. The two polar views show the largest amplitudes and the variation of the northern hemisphere is opposite to the southern one. The signals from the equatorial views are also detectable in the disk-integrated data. However, the amplitude is one to two orders of magnitudes smaller depending on the molecular absorption feature. The full, non-averaged seasonality plots for the individual viewing geometries are shown in Appendices 12–15.

**Table 3**
Atmospheric Seasonality for the Different Viewing Geometries and Molecular Absorption Features

| Viewing Geometry | Mol. Absorption Feature | Annual Mean EW (nm) | Seasonal Range (nm) | Seasonal Variation [%] |
|---|---|---|---|---|
| NP | $O_3$ | 358.88 | 13.36 | 3.72 |
| | $CH_4$ | 51.10 | 2.42 | 4.74 |
| | $CO_2$ | 266.90 | 23.08 | 8.64 |
| | $N_2O$ | 30.44 | 0.36 | 1.18 |
| SP | $O_3$ | 353.59 | 8.04 | 2.27 |
| | $CH_4$ | 50.21 | 2.20 | 4.38 |
| | $CO_2$ | 262.36 | 41.34 | 15.76 |
| | $N_2O$ | 30.17 | 0.16 | 0.53 |
| EqA | $O_3$ | 351.59 | 5.30 | 1.51 |
| | $CH_4$ | 50.65 | 0.88 | 1.74 |
| | $CO_2$ | 265.22 | 4.42 | 1.67 |
| | $N_2O$ | 30.59 | 0.18 | 0.58 |
| EqP | $O_3$ | 350.84 | 4.28 | 1.22 |
| | $CH_4$ | 49.84 | 0.82 | 1.64 |
| | $CO_2$ | 265.36 | 6.84 | 2.58 |
| | $N_2O$ | 30.37 | 0.10 | 0.32 |

**Note.** Besides the annual mean EW and the range of the oscillation, the table states the seasonal variation in the last column where the relative change of the range is compared to the annual mean EW.

of carbon monoxide (CO) is a particularly compelling biosignature since the combination of the two represents carbon in its most reduced and most oxidized forms, which is hard to explain without life (Krissansen-Totten et al. 2018). Also, methane's coexistence with ozone is possibly the strongest biosignature. Several sinks exist for methane, yet the most dominant one involves the oxidation of $CH_4$ with the hydroxyl (OH) or chlorine (Cl) radicals (e.g., Grenfell et al. 2007, Figure 4). Secondary sinks include, for example, photolysis and dry deposition (Grenfell 2018; Schwieterman et al. 2018). Although in terrestrial exoplanet atmospheres, $CH_4$ has a short photochemical lifetime and requires substantial replenishment fluxes in order to accumulate detectable abundances, atmospheric $CH_4$ on Earth is rather unreactive in the troposphere, residing there for $9.1 \pm 0.9$ yr. Due to its relatively long lifetime compared to the length of the seasonal cycle, its sources being predominantly located near the surface, and the vertical atmospheric transport evening out regional methane differences, $CH_4$ is considered a well-mixed gas in the troposphere. Yet, satellite measurements also detected methane at stratospheric levels, indicating that winds transport plumes of gas considerable distances from their sources.

The annual strength of the $CH_4$ absorption feature deviates by 1.08% for the different viewing geometries. The amplitudes of the atmospheric seasonality of methane for the two polar regions are 3 times larger than for the equatorial viewing geometries and their relative annual *growth* in EW is 4.74% and 4.38% for NP and SP, respectively. Comparing the two hemispheres to each other, the northern hemisphere pole-on view shows a 9% larger seasonality in CH$_4$ than SP. Between the two equatorial views, the EqA view, shows a 7% larger amplitude than the ocean-dominated view EqP, leading to a relative annual growth in EW of 1.74% and 1.65% for EqA and EqP, respectively.

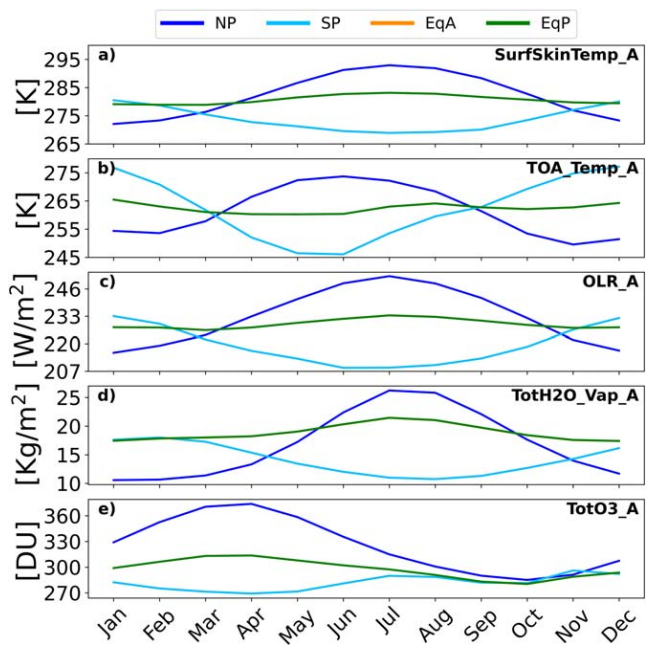

**Figure 10.** Panels (a)–(e) show 4 yr monthly averages of the surface skin temperature, top of atmosphere (TOA) temperature, outgoing long-wave radiation (OLR), total integrated column water vapor burden, and the total integrated column ozone burden, respectively. The data originate from level 3 Earth observation data (AIRS3STD) and was disk averaged for the four viewing geometries for dayside only. AIRS3STD provides thermodynamic and trace gas parameters such as the skin temperature for land and sea surfaces, which result from retrievals on the standard pressure levels roughly matching instrument vertical resolution. Due to small differences between the two equatorial views EqA and EqP, the data points (orange) of the former are not visible.

Inspecting the EW time series of NP and SP (see Figures 12 and 13 for a more detailed view), we can identify atmospheric processes that were outlined in Section 2.2 Olson et al. (2018b). The seasonal CH$_4$ cycle observed for the southern hemisphere is primarily photochemical. Due to the enhanced evaporation during summer time, the increased concentration of tropospheric H$_2$O leads to greater photochemical production of OH radicals that oxidize methane at an accelerated rate (see panels (a)–(d) in Figure 10). The opposite effect occurs during winter months, where the cold temperatures mute the destruction of CH$_4$. The seasonality of methane in the northern hemisphere, on the other hand, shows a biological cycle that is out of phase with the photochemical cycle, leading to a second peak in its time series (Khalil & Rasmussen 1983). Hence, the observed temporal oscillation for methane in the top right panel of Figure 9 is dominantly controlled by the photochemical cycle. For the northern hemisphere, the highest abundance of atmospheric methane occurs in late fall and early spring and the minimum in summer and winter, where for the latter the annual minimum concentration is reached. Thus, the abundance of CH$_4$ in Earth's atmosphere is more strongly correlated with the surface temperature and the resulting evaporation of water as well as the solar zenith angle that controls the production of hydroxyl radicals than the instantaneous release from biogenic sources (Khalil & Rasmussen 1983). The abiotic response of methane to the seasonality of tropospheric H$_2$O is particularly interesting as its oscillation indicates the presence of a large surface water reservoir in liquid state and in combination with its long lifetime against loss methane is a tracer of dynamical motions in the lower atmosphere.

Carbon dioxide (CO$_2$) and its seasonal variation in atmospheric composition due to photosynthetic activity is possibly the most well-documented and mechanistically understood biologically modulated occurrence of all the key spectrally active gases (e.g., Keeling 1960). With the growth of vegetation mass in spring through summer, the rate of photosynthesis increases and the atmospheric abundance of CO$_2$ decreases as the carbon is fixed into organic matter. During fall and winter the rate of photosynthesis, and thus the consumption of CO$_2$, slows as plant matter decays, leading to a rise in its atmospheric concentration. Since the photosynthetic activity is the cause of seasonal fluctuation in CO$_2$, regions with a higher landmass fraction experience a larger magnitude due to the vegetative cover that drives the seasonal cycle. Both the northern hemisphere continents and the tropics include the majority of land plants. Yet, near the equator, the seasonal variations in temperature are less pronounced than at higher latitudes where the seasonal changes in temperature results in large seasonal CO$_2$ variations. Furthermore, photosynthesis also occurs in the oceans by plankton, algae, and some cyanobacteria; however, the sea-air gas exchange flux of CO$_2$ is ~0.1% of the total natural contribution to the carbon dioxide budget (Canadell et al. 2021). Hence, for Earth, there is a latitudinal and hemispherical difference in the magnitude of the CO$_2$ fluctuation, where the northern hemisphere, due to its larger continental mass and larger seasonal temperature fluctuations, shows a greater amplitude, overall. The oscillation ranges from ~3 ppm near the equator to ~10–20 ppm at higher latitudes in the northern hemisphere (Keeling 1996).

In order to analyze carbon dioxide's seasonal change in absorption strength, we have investigated the CO$_2$ absorption feature centered at 4.3 $\mu$m whose emission originates in altitudes of up to ~10 km. The global mean EW for this CO$_2$ feature deviates by 0.71% for the four viewing geometries. In terms of amplitudes, however, they differ by 90% where the largest difference is due to the two hemispheres that show a seasonal range of 23.08 and 41.34 nm for NP and SP, respectively. Although the northern hemisphere shows a 1.7% larger annual mean EW than the southern hemisphere, unexpectedly, SP shows a 79% larger amplitude than the northern hemisphere view NP, resulting in a seasonal variation of 15.76%. SP's seasonal variation is therefore the largest one across the observing geometries and target molecules, followed by NP's variation of 8.65% for CO$_2$.

The annual mean EW of the two equatorial views differs by 0.14 nm, yet the Pacific-dominated observing geometry, EqP, shows a 1.5 times larger seasonal variation than the Africa-centered view EqA. A possible explanation could be due to the fact that the tropical oceans and the high latitude oceans, particularly the southern ocean, contribute the most to the global mean interannual variability. Observation-based $p$CO$_2$ flux measurements show that emissions of natural CO$_2$ occur mostly in the tropics and high latitude southern oceans, whereas strong ocean $CO_2$ sink regions are found in the midlatitudes associated with the cooling of poleward flowing subtropical surface waters as well as equatorward flowing subpolar surface waters (e.g., DeVries et al. 2019; Canadell et al. 2021; Long et al. 2021). Comparing the two equatorial views, EqP contains a very prominent $CO_2$ outgassing region (also a global maxima), which extends from the north of South America along the equator to Indonesia (e.g., Figure 5.9 in Canadell et al. 2021) with a net air-sea flux of 3 times and higher the value of the corresponding region at similar latitudes in EqA.

Furthermore, nLTE effects make a large contribution to the flux of the core of the 4.3 $\mu$m $CO_2$ band, which might be impacting the seasonal measurements as well (e.g., DeSouza-Machado et al. 2007; López-Valverde et al. 2011).

Nitrous oxide ($N_2O$) is a strong greenhouse gas primarily produced by Earth's biosphere as a by-product during the remineralization of organic matter via processes of nitrification and denitrification on land and ocean (e.g., Tian et al. 2020). Its tropospheric abundance in modern Earth's atmosphere is $332.1 \pm 0.4$ ppb with a mean atmospheric lifetime of $116 \pm 9$ yr on Earth. The dominant sinks of $N_2O$ involve photolysis and oxidation by electronically excited oxygen atoms, $O(^1D)$, in the stratosphere (e.g., Grenfell et al. 2007, Figure 4), resulting in an annual loss of $\sim 13.1$ TgN yr$^{-1}$ (Canadell et al. 2021, and references therein). It also contributes to the destruction of stratospheric ozone and its emission has currently the largest ozone depletion potential of all ozone-depleting substances (Lessin et al. 2020). Variability in its atmospheric abundance is affected by the net $N_2O$ sources on the ground and the photochemical destruction in the stratosphere. Yet, the production is highly sensitive to environmental conditions such as temperature, ph, and oxygen concentrations, among many others, causing strong variability of $N_2O$ emissions in time and space. Due to the fact that abiotic sources of $N_2O$ are two orders of magnitude lower than the biological sources and its potential spectral detectability with several absorption lines across the MIR with significant bands centered at 3.7, 4.5, 7.8, 8.6, and 17 $\mu$m, nitrous oxide has been proposed as a strong biosignature (Sagan et al. 1993; Segura et al. 2005; Catling et al. 2018; Schwieterman et al. 2018; Canadell et al. 2021). For Earth's abundances of $N_2O$, however, most of these bands are weak and/or overlap with $CO_2$, $H_2O$, and $CH_4$, requiring a high spectral resolution power to identify individual lines in order to differentiate from overlapping gas absorption features. Yet, in terms of exoplanets, studies show that these features can become more significant for planets in weak-UV environments where $N_2O$ builds up (e.g., Segura et al. 2005; Grenfell & Gebauer 2014; Rugheimer et al. 2015).

From the fourth panel of Figure 9, we deduce that the Africa-centered view, EqA, shows the highest mean EW (30.59 nm) followed by NP (30.44 nm), EqP (30.37 nm), and SP (30.17 nm), in descending order. With a global equivalent width average of 30.39 nm, $N_2O$ shows the least variation of all the target molecules between the different observing geometries, which is essentially due to its long photochemical lifetime and solubility. Its deviation is 0.57% ($\sigma = 0.17$ nm). Yet, the largest seasonal range is measured for NP (0.36 nm), which is a factor of 2 larger than the amplitudes of SP (0.16 nm) and EqA (0.18 nm) and a factor of 3 larger than the ocean-dominated view EqP (0.10 nm). The resulting annual change in strength, compared to the annual mean EW, is 1.18%, 0.53%, 0.59%, and 0.33% for NP, SP, EqA, and EqP, respectively. Thus, the northern hemisphere pole-on view, NP, shows a seasonal variation of $N_2O$ that is twice as large as the southern hemisphere pole-on view, SP, and the Africa-centered view, EqA, and a factor of 3.7 times larger than the ocean-dominated view EqP. In terms of range, the two hemispheres differ by 56% and the two equatorial views by 44%.

The oscillations from observation geometries that include emission from the northern hemisphere (NP, EqA, and EqP) show a complex structure containing two to three peaks in their seasonality (see Figures 12 and 15 in Appendix B). It is particularly visible for the EqA observation geometry. Whether this is the effect of a biological cycle that is out of phase with the photochemical cycle in combination with the increased biological rate due to the warmer climate and higher temperatures during summer requires further research. Yet, according to the Intergovernmental Panel on Climate Change (IPCC), oceans contribute 21.2% to the $N_2O$ emission to the atmosphere. For the northern hemisphere, the air-sea fluxes are the highest between late spring to summer, where there is a difference between the emission peak of near coastal and offshore water regions. The latter's emission peaks in late fall. Hence, physical constraints regulate seasonal patterns of $N_2O$ fluxes. The duration and intensity of water column stratification as well as biological productivity define the timing, abundance, rate, and vertical position of $N_2O$ released into the atmosphere (Lessin et al. 2020; Canadell et al. 2021, and references therein).

## 4. Discussion

Rotational and seasonal variations of Earth's spectrum and their influence on the detectability of spectral signatures of habitability and biosignatures have been investigated before, both based on observational data and simulations as well as in the reflected light and thermal emission (e.g., Ford et al. 2001; Cowan et al. 2009; Hearty et al. 2009; Fujii et al. 2011; Livengood et al. 2011; Robinson 2011; Gómez-Leal et al. 2012; Robinson et al. 2014; Olson et al. 2018b; Jiang et al. 2018; Mettler et al. 2020).

We have adopted some aspects of the method for disk-integrating the data from previous studies that investigated Earth's emission in the MIR, e.g., Tinetti et al. (2006a) or Hearty et al. (2009). Yet, our data set differs especially in terms of time baseline, temporal resolution, and the effective number of disk-integrated spectra based on satellite observation data. Both previously mentioned studies took advantage of Earth observation data collected by the AIRS instrument, which they either used as input data for designing their model, like the former, or studied 1 day per month for a 1 yr/full orbit, as in the latter. Furthermore, the work by Gómez-Leal et al. (2012) presents a data set of photometric time series with a resolution of 3 hr and a baseline of 22 yr. Hence, providing a data set with a greater baseline and better time resolution. However, for their study they derived the disk-integrated photometric signal of the Earth using top-of-atmosphere all-sky upward long-wave fluxes that are integrated over a 4–50 $\mu$m wavelength interval, thus, not providing spectral information. Therefore, our exclusive data set comprising 2690 disk-integrated MIR thermal emission spectra with a spectral resolution of 3.75–15.4 $\mu$m ($R \sim 1200$) derived from remote sensing observations for four full-disk observing geometries over four consecutive years (2016–2019) at high temporal resolution

 

presented in this work is unique and allows us to study the time-variable thermal emission and the atmospheric seasonality of Earth from afar in great detail.

Despite differences in the nature of data sets, results, and trends associated with planetary obliquity as well as radiance levels for comparable regions are in agreement with these studies. In further agreement, at the investigated resolution of $R \approx 1200$, molecular lines for example of methane, carbon dioxide, ozone, water, and nitrous oxide are detectable in the disk-integrated spectra, independent of viewing geometry, time of observation, or cloud fraction. Furthermore, our results of disk-integrated spectra reflect the following two statements of Hearty et al. (2009): (1) the $O_3$ and $CO_2$ features can appear either in emission or absorption due to temperature inversions in Earth's atmosphere and (2) the 15 $\mu$m $CO_2$ feature is less sensitive to the day-night difference than the 4.3 $\mu$m $CO_2$ absorption feature.

### 4.1. Complementarity of UV–Vis NIR and MIR Observations

A detailed characterization of the planet's surface, atmosphere, and potential habitability would require the combination of reflected light and thermal emission observations. The MIR thermal emission spectrum of Earth from afar and its time variability contain a wealth of information about the atmospheric and surface environment. Observing in the MIR enables us to characterize the thermal structure of exoplanetary atmospheres and provides additional information on the molecular composition. In general, more molecules have strong absorption bands in the MIR and even at relatively low spectral resolutions, absorption features of key greenhouse gases and/or bioindicators such as $O_3$, $CH_4$, $CO_2$, $N_2O$, and $H_2O$ can be detected (e.g., Christensen & Pearl 1997; Hearty et al. 2009). Furthermore, compared to visible wavelengths, the IR is less affected by the presence of hazes and clouds which is a major challenge for the former (e.g., Kitzmann et al. 2011; Fauchez et al. 2019). Moreover, pressure-induced absorption features can indicate bulk atmospheric composition and pressure (e.g., Schwieterman et al. 2015). The observed flux in the MIR can be used for constraining the radius of a terrestrial exoplanet, which is much more difficult in the reflected light. With the planetary radius determined from IR observations, broadband photometric observations can also constrain the planetary albedo, which is necessary to comprehend the planetary energy balance (Pallé et al. 2003). Furthermore, broadband photometric and spectroscopic observations can reveal habitable environments as well as provide information about atmospheric and surface environments. For example, from reflected light observations a planetary color could be used to identify exo-Earth candidates (Traub 2003; Fujii et al. 2010), the effect of specular reflectance on a planetary phase curve could reveal surface oceans (Robinson et al. 2010), the vegetation *red edge* in the reflectance spectra would be a strong surface biosignature of land vegetation (e.g., Des Marais et al. 2002; Schwieterman et al. 2018), and the absorption features in the reflected light are suitable for abundance determination as they are not affected by the thermal structure of the atmosphere (e.g., Drossart et al. 1993; Livengood et al. 2011).

With regard to seasonality, however, observing in the MIR offers advantages over observing the reflected light of a planet. In this case, the MIR will not be negatively impacted by the lack of illumination of the winter hemisphere over the course of the orbit (e.g., see Figure 5 in Olson et al. 2018b). Thus, combining the information from both, the reflected and thermal emission, observations could break some day-night degeneracies. In addition, more information could be gained about tropospheric water transport and cloud variability from seasonal changes in depths of $H_2O$ bands, which could be helpful for retrievals in the MIR. Especially since it remains unclear how patchy clouds would influence our ability to retrieve and understand the atmosphere and surface of an Earth twin from thermal emission observations.

### 4.2. Clouds and Their Influence on the Thermal Emission Spectra

Although we did not investigate and include additional information on cloud fraction and their thermodynamical phase properties in our study, roughly 67% of Earth is covered by clouds at all times (King et al. 2013). Hence, cloud seasonality and their effect on the MIR thermal emission and the detectability of spectral features, as for example shown in Des Marais et al. (2002), Tinetti et al. (2006a, 2006b), and Hearty et al. (2009), is imprinted in our derived spectra, and thus, part of our results.

Gas oscillations in Earth's modern biosphere vary in the order of 1%–3% for $CO_2$ and $CH_4$ (e.g., Schwieterman et al. 2018), indicating that our values of seasonal variations in EWs for NP and SP of 2.37% and 2.19% for Earth and 4.32% and 7.88% for $CO_2$ (at 4.3 $\mu$m) are affected most probably by varying cloud covers. This is supported by the fact that the origin of the layer of emission for the $CH_4$, $CO_2$, and $N_2O$ features studied in this work lies near the middle to upper troposphere, i.e., about 200–300 hPa in the tropics and 400–500 hPa in the polar regions for methane, 300–900 hPa for $N_2O$ and 300–1000 hPa for $CO_2$ at the aforementioned wavelengths (e.g., Goody & Hu 2003, Figure 4). However, differences in the $H_2O$ continuum are also impacting $CH_4$ and non-LTE effects are impacting $CO_2$.

Only the ozone feature centered at 9.6 $\mu$m lies well above the cloud covers as these wavelengths probe layers from the upper troposphere and lower to the mid stratosphere at 10 to 30–50 km in altitude. Although the trend of the annual seasonality in EWs followed the trend of the abundance for $O_3$, we calculated a seasonal variation between 0.61% and 1.86% depending on the viewing geometry, which differs from the seasonal variability in abundance whose amplitude varies by 10%–15% (at 25–35 km) in the total column value of ~10 ppm (e.g., Schneider et al. 2005). The ozone absorption feature is saturated (like the $CO_2$ feature at 15 $\mu$m); thus, it will not vary linearly with abundance, and its measured spectral variability may be smaller, which could explain the difference. We plan to investigate in future studies how cloud fraction, cloud seasonality, and their thermodynamical phase properties affect Earth's thermal emission spectra as well as the detection and the result of atmospheric seasonality.

### 4.3. Drivers of Seasonal Variability for Terrestrial Planets

The last 25 yr of detection and characterization of exoplanets revealed a vast diversity of planets regarding their masses, sizes, and orbits (e.g., Batalha 2014; Burke et al. 2015; Paradise et al. 2022). It is expected that this diversity also extends to their obliquity and atmospheric mass and composition. Hence, terrestrial exoplanets could display seasonality that is very different from that of Earth or any other solar system planet. For example, the seasonal signal could be boosted if photochemical lifetimes were shorter and/or saturated bands were not saturated. Regarding the latter, there is a trade-off between the abundance of a gas and its variability which is really key for potentially detecting seasonal changes. Eccentric planets may have better-characterized seasonality as the competing effects of admixed hemispheres will not exist. With increasing orbital obliquity, the seasonal contrast increases as the ice and vegetation cover would vary. A moderately high obliquity promotes increased photosynthetic activity and associated oxygen flux to the atmosphere. In general, the biological activity will be increased in such a scenario leading to heightened variations in biosignature gases over more intense seasonal cycles, making life perhaps easier to detect (Barnett & Olson 2022). Yet, as the detectability of seasonality depends on both the magnitude of the biogenic signal and the extent to which observation conditions mute that signal, the detectability of seasonality as a biosignature is likely optimized at intermediate obliquity that is sufficient to produce a large-magnitude signal but not so large as to preclude viewing of the winter hemisphere (Olson et al. 2018b).

### 4.4. Anthropogenic Contribution to the Data

Using data from Earth-orbiting satellites to study its thermal emission and, especially, the seasonality of its bioindicators in order to apply it to exoplanets has some limitations due to the anthropogenic contribution. Specifically, the abundance of biosignatures like $CH_4$ or $N_2O$, which are both more powerful greenhouse gases than $CO_2$ with their high radiative forcing capabilities, are strongly affected by human activities. For example, according to Canadell et al. (2021) the human contribution and perturbation of the natural nitrogen cycle through the use of synthetic fertilizers and manure, as well as nitrogen deposition resulting from land-based agriculture and fossil fuel burning has been the largest driver of the increase in atmospheric $N_2O$ of $31.0 \pm 0.5$ ppb (10%) between 1980 and 2019. The average annual tropospheric growth rate in the most recent decade of 2010–2019, in which our consecutive year sample of 2016–2019 lies, was $0.85 \pm 0.03$ ppb yr$^{-1}$. The growth rate of methane and carbon dioxide was even larger in the same decade. Hence, the anthropogenic factor needs to be accounted for, for longer time baseline studies. In the case of our 4 yr sample, a slightly increasing trend is visible for $N_2O$ and $CH_4$ in the EW plots, specifically, for NP, SP, and EqP (see Figures 12, 13, and 15 in Appendix B).

## 5. Summary and Conclusions

In this work, we presented our exclusive data set comprising 2690 disk-integrated MIR thermal emission spectra derived from remote sensing observations for four different viewing geometries at a high temporal resolution over a time baseline of 4 yr. Using this data set, we have investigated how Earth's MIR spectral appearance changes as a function of viewing geometry, seasons, and phase angles and quantified the atmospheric seasonality of different bioindicators. We found a representative, disk-integrated thermal emission spectrum of Earth does not exist. Instead, there is significant seasonal variability in Earth's thermal emission spectrum, and the strength of biosignature absorption features depends strongly on both season and viewing geometry.

Earth's appearance from afar is dominated on large scales by oceans, deserts, ice as well as vegetation and clouds. The contribution of these different surface types (and climate zones) to the latitudinally disk-integrated signal and the annual variability depends on their abundance fraction and distribution as well as on their thermal properties. Thus, landmass-dominated Earth views did not only show higher flux readings but larger annual variabilities over one full orbit than ocean-dominated views. Specifically, the northern hemisphere pole-on view (NP) and the Africa-centered equatorial view (EqA) showed annual variabilities of 33% and 22% at Earth's peak wavelength at $\approx$10.2 $\mu$m, respectively. On the other hand, viewing geometries with a high sea fraction such as the southern hemisphere pole-on (SP) and the Pacific-centered, equatorial view (EqP) show smaller annual variabilities due to the large thermal inertia of oceans. In this specific case, both viewing geometries displayed a similar variability of $\sim$11% at 10.2 $\mu$m, therefore, varying by a factor of 3 and 2 less compared to that of NP and EqA, respectively. Furthermore, concerning near pole-on observations that will probe only one hemisphere, we found that for Earth the annual variability of the two hemispheres differs by a factor of 3 across all the spectral windows evaluated in this study. Out of the four viewing geometries, the ocean-dominated view EqP showed the least variability in terms of annual, seasonal, and diurnal variations.

Due to the disk-integrated data and the annual variability of the thermal emission spectrum, the data shows a strong degeneracy. Differentiating between the two hemispheres is especially challenging during the vernal equinox since the spectra of NP and SP overlap in all three spectral windows. Furthermore, the large annual variability of NP covers the flux ranges of EqP and SP, indicating the difficulties of uniquely interpreting and characterizing planetary characteristics from disk-integrated data. In the case of Earth, the only viewing geometry that remained separated flux-wise from other Earth views at Earth's peaking wavelength in the MIR, is the Africa-centered equatorial view EqA, despite, showing the second largest annual variability. Thus, without sufficient knowledge about a planet's orbital parameters and obliquity, even for Earth, interpreting the space and time-averaged data based on single-epoch measurements is quite challenging. Hence, multi-epoch measurements and the resulting time-dependent signals may be required to help break the degeneracy in the thermal emission spectra in order to fully characterize a planetary environment.

We quantified the amplitudes and seasonal variation in absorption strength by calculating the EWs of absorption features of biosignatures imprinted in the disk-integrated spectra. The detected variability is reduced for certain viewing geometries, i.e., from pole-on toward equatorial views. In general, the northern hemisphere pole-on observation geometry showed larger amplitudes of atmospheric seasonality for the investigated bioindicators followed by SP and the two equatorial views. Except for carbon dioxide, whose seasonal variation was found to be the largest among all targets, where seasonal variations of 15.76% and 8.65% for SP and NP were derived, respectively. In Earth's modern biosphere, gas oscillations vary typically in the order of 1%–3% for $CO_2$ and $CH_4$; thus, the seasonal variation in EW is affected by varying cloud covers that induce additional variation plus potentially non-LTE effects for $CO_2$. This is further supported by the fact that the layer of emission lies in the middle to upper troposphere. However, differences in the $H_2O$ continuum are also impacting $CH_4$ and non-LTE effects are impacting $CO_2$. Future work is required to investigate how cloud fraction, cloud seasonality, and their thermodynamical phase properties affect the detection and result of atmospheric seasonality.

Using Earth as our test bed, we learned that a planet and its characteristics cannot be described by a single thermal emission spectrum but multi-epoch measurements, preferably in both reflected light and thermal emission, are required. Its spectral appearance in terms of flux levels and strength of spectral features changes continuously along the orbit due to processes driven by obliquity and eccentricity which impact nearly every constituent in the atmosphere and, if present, biosphere. Moreover, we found a strong spectral degeneracy with respect to viewing geometry due to variable blending of time-variable thermal emission from hemispheres with opposing seasonal signals in the disk-integrated views.

This complexity makes remote characterization of planetary environments very challenging. Yet, the degeneracy is in favor of MIR thermal emission observations since irrespective of when a planet would be observed the overall variation (in flux or absorption bands) is typically ≲10% for a given viewing angle. Although disentangling these variations from the noise in future observations is nontrivial, we find that our result is relatively insensitive to diurnal or seasonal effects, unlike in the case of reflected light measurements. We, therefore, conclude that observing exoplanets with thermal emission could provide unique and complementary information that is necessary for the characterization of terrestrial planets around other stars.

We thank an anonymous referee for the valuable comments. This work has been carried out within the framework of the National Center of Competence in Research PlanetS supported by the Swiss National Science Foundation. The authors acknowledge the financial support of the SNSF. S.L.O and E. W.S. gratefully acknowledge support from the NASA Exobiology Program via grant 80NSSC20K1437 and the NASA Interdisciplinary Consortia for Astrobiology Research (ICAR) program via the Alternative Earths team with funding issued under grant No. 80NSSC21K0594. J.N.M. carried out the analyses, created the figures, and wrote the bulk of the manuscript. S.P.Q. initiated the project. S.P.Q. and R.H. guided the project and wrote part of the manuscript. All authors discussed the results and commented on the manuscript.

## Appendix A
## Disk-integrated Thermal Emission Spectra for NP, SP, and EqC

Figure 11 in Appendix A, compares the disk-integrated spectra of the two hemispheres to the combined equatorial view EqC.

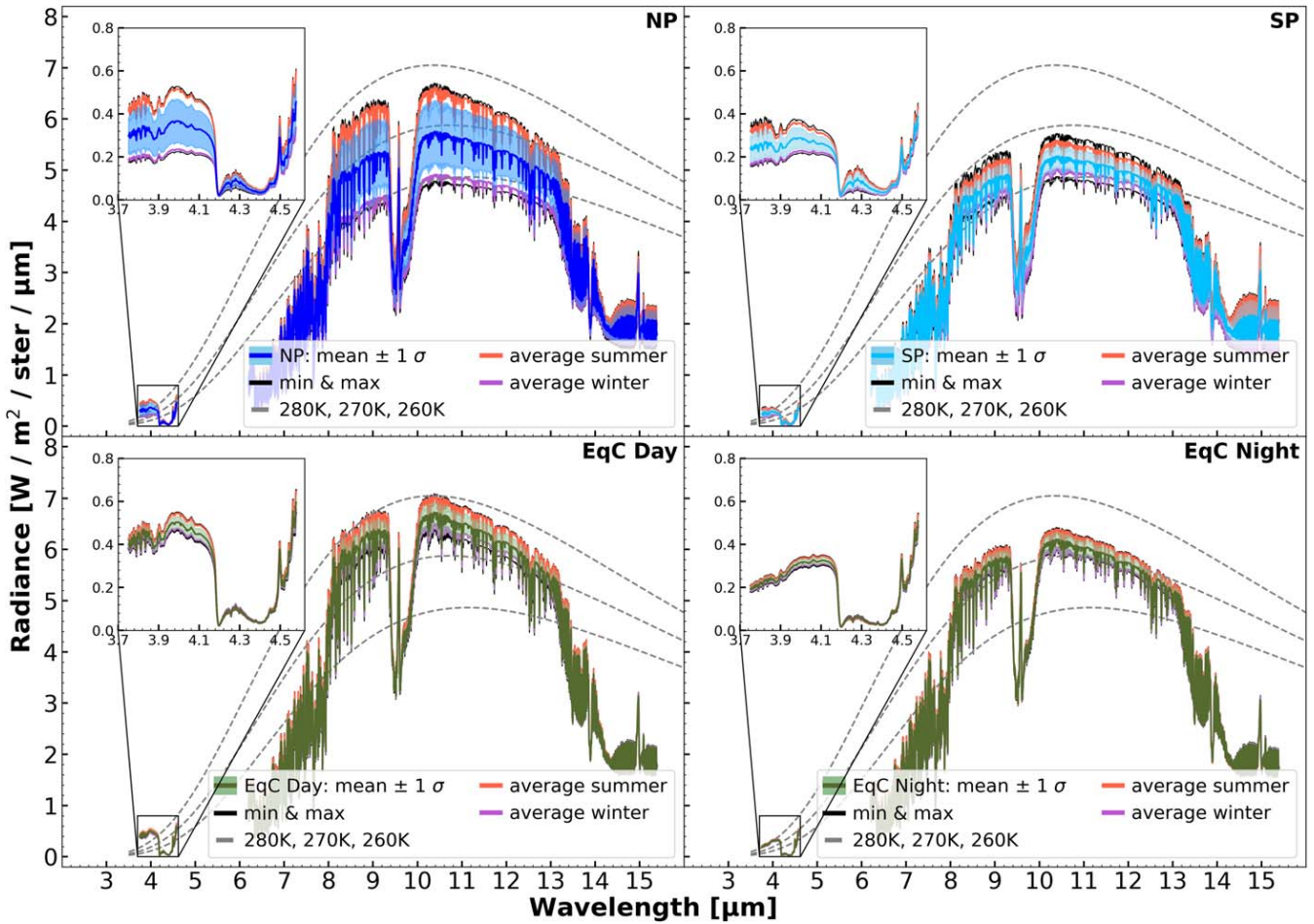

**Figure 11.** A comparison of the disk-integrated thermal emission spectrum for three different observing geometries (NP, SP, and EqC). The mean represents the annual spectrum averaged over 4 yr. The shaded area corresponds to the standard deviation of all measurements for that particular observing geometry. Summer and winter were defined as the months with the highest and lowest flux levels at 10 $\mu$m, respectively. For the northern hemisphere, this turned out to be July and January, and for the South Pole vice versa.

## Appendix B
## Atmospheric Seasonality Figures

In Appendix B, Figures 12–15 show the full temporal variations in abundance for each viewing geometry over four years.

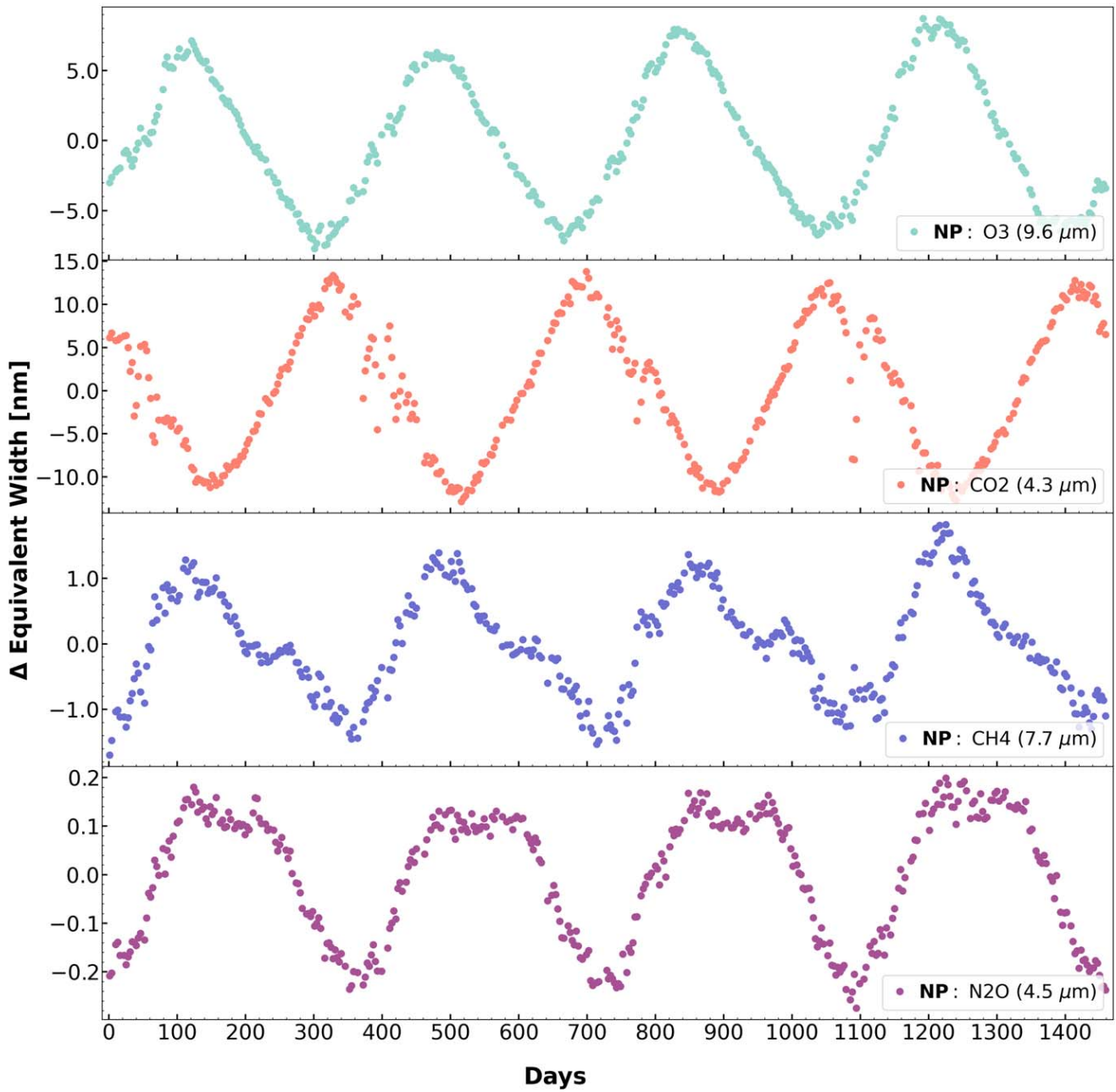

**Figure 12.** Viewing geometry: NP.

## Atmospheric Seasonality - O3, CO2, CH4, N2O

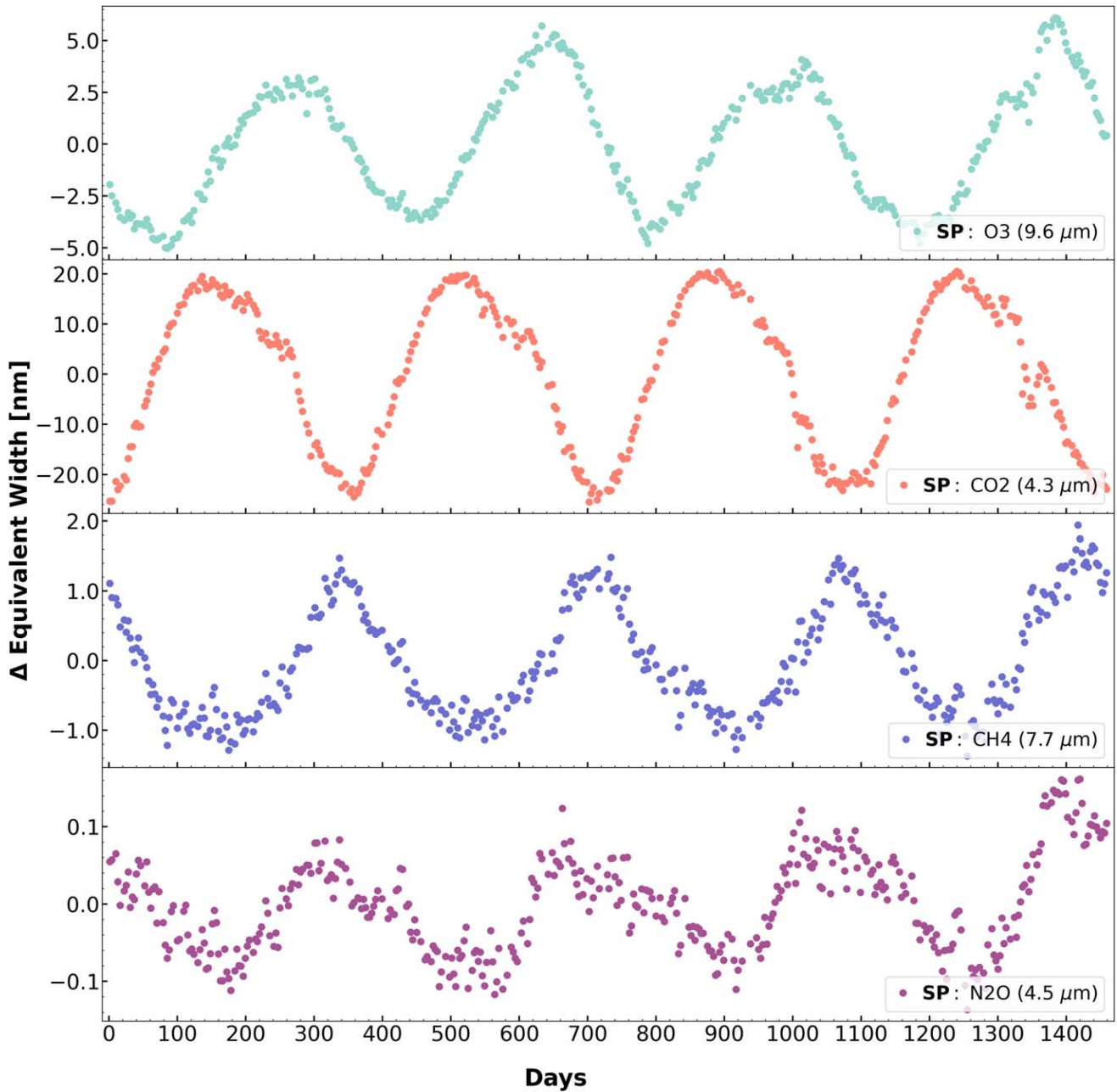

**Figure 13.** Viewing geometry: SP.

## Atmospheric Seasonality - O3, CO2, CH4, N2O

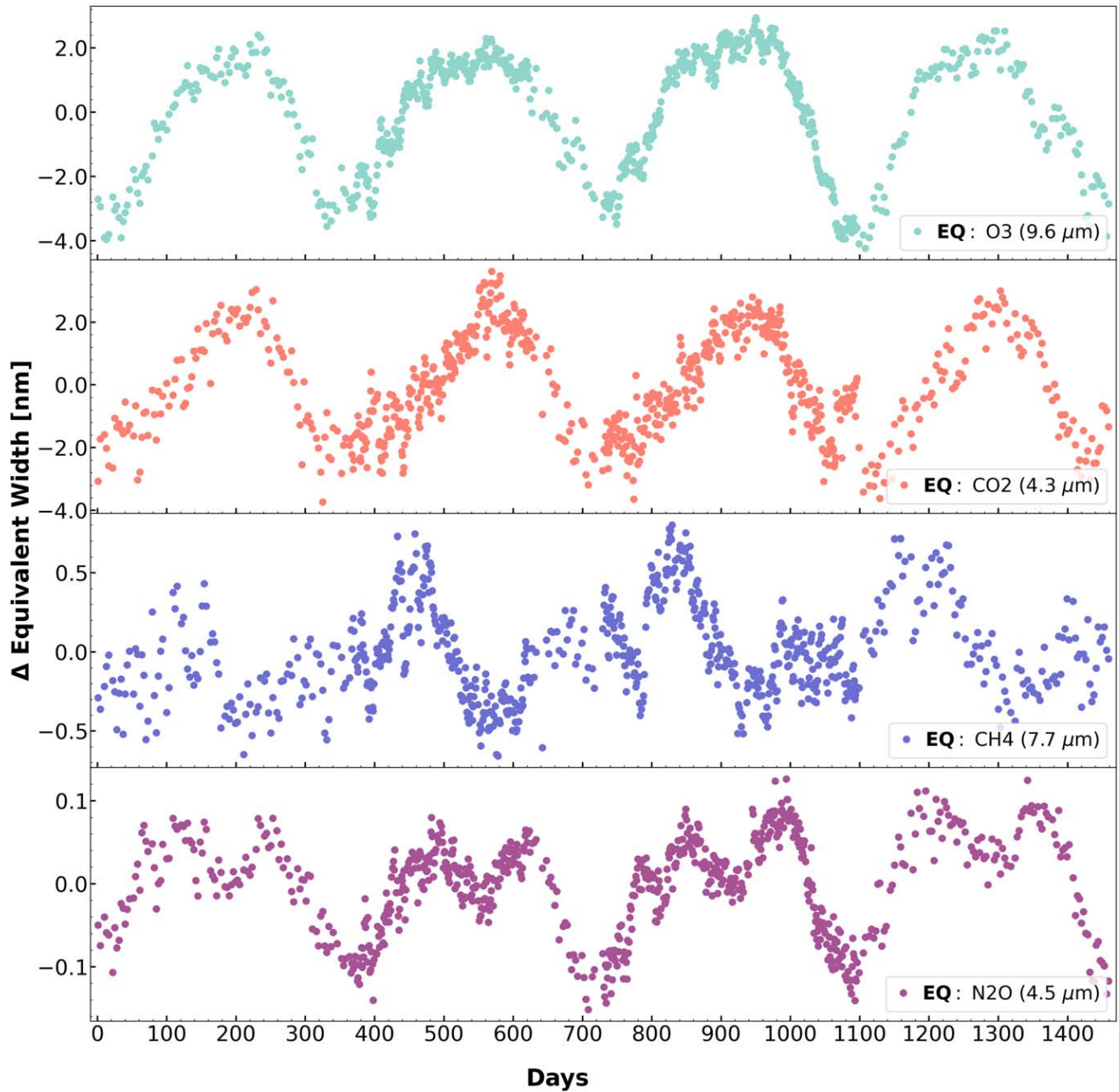

**Figure 14.** Viewing geometry: EqA.

## Atmospheric Seasonality - O3, CO2, CH4, N2O

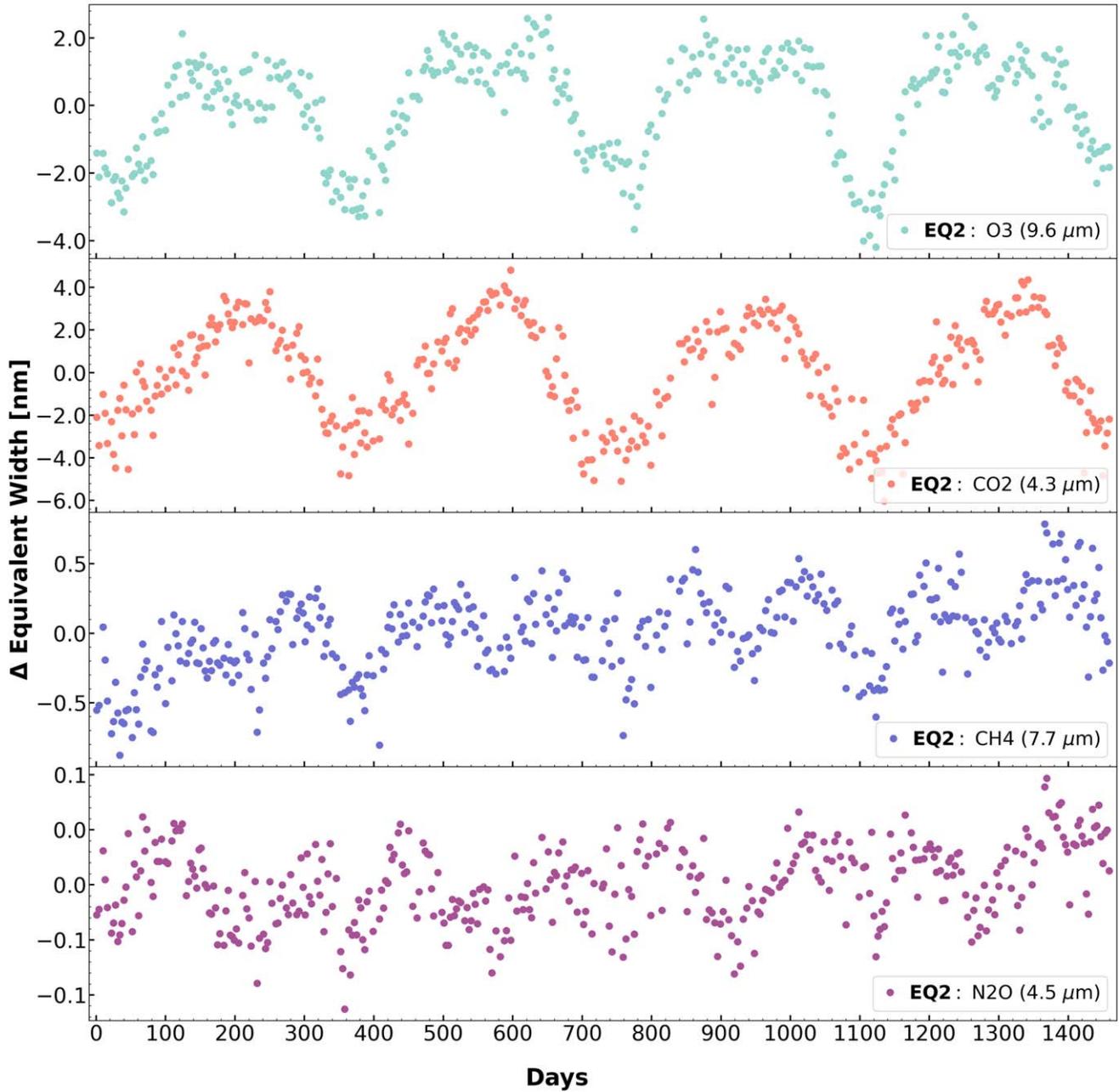

**Figure 15.** Viewing geometry: EqP.

## ORCID iDs

Jean-Noël Mettler  https://orcid.org/0000-0002-8653-0226
Sascha P. Quanz  https://orcid.org/0000-0003-3829-7412
Ravit Helled  https://orcid.org/0000-0001-5555-2652
Stephanie L. Olson  https://orcid.org/0000-0002-3249-6739
Edward W. Schwieterman  https://orcid.org/0000-0002-2949-2163